\documentclass[aip, jcp, reprint]{revtex4-1}

\usepackage{xspace}
\usepackage{graphicx}
\usepackage{subcaption}
\usepackage{circledsteps}

\newcommand{\model} {{\sf Model}\xspace}
\renewcommand{\models}{{\sf Models}\xspace}
\newcommand{\test}  {{\sf Test}\xspace}
\newcommand{\tests} {{\sf Tests}\xspace}
\newcommand{\modeldriver} {{\sf Model~Driver}\xspace}
\newcommand{\modeldrivers}{{\sf Model~Drivers}\xspace}
\newcommand{\testdriver}  {{\sf Test~Driver}\xspace}
\newcommand{\testdrivers}  {{\sf Test~Drivers}\xspace}

\newcommand{\verificationcheck} {{\sf Verification~Check}\xspace}
\newcommand{\verificationchecks} {{\sf Verification~Checks}\xspace}

\newcommand{\error} {{\sf Error}\xspace}

\newcommand{\verificationresult} {{\sf Verification~Result}\xspace}
\newcommand{\verificationresults} {{\sf Verification~Results}\xspace}
\newcommand{\testresult} {{\sf Test~Result}\xspace}
\newcommand{\testresults} {{\sf Test~Results}\xspace}

\begin{document}

\title[The OpenKIM Processing Pipeline]{The OpenKIM Processing Pipeline: A Cloud-Based Automatic Materials Property Computation Engine}

\author{D.~S.~Karls}
  \affiliation{Department of Aerospace Engineering and Mechanics, University of
Minnesota, Minneapolis, MN 55455, USA}

\author{M.~Bierbaum}
  \affiliation{Department of Information Science, Cornell University, Ithaca, NY 14850, USA}

\author{A.~A.~Alemi}
  \affiliation{Google Research, Mountain View, CA, USA}

\author{R.~S.~Elliott}
  \affiliation{Department of Aerospace Engineering and Mechanics, University of
Minnesota,  Minneapolis, MN 55455, USA}

\author{J.~P.~Sethna}
  \affiliation{Department of Physics, Cornell University, Ithaca, NY 14850, USA}

\author{E.~B.~Tadmor}
  \affiliation{Department of Aerospace Engineering and Mechanics, University of
Minnesota,  Minneapolis, MN 55455, USA}
  \email{tadmor@aem.umn.edu}

\date{\today}

\begin{abstract}
The Open Knowledgebase of Interatomic Models (OpenKIM) project is a framework intended to facilitate access to standardized implementations of interatomic models for molecular simulations along with computational protocols to
evaluate them. These protocols includes tests to compute materials properties
predicted by models and verification checks to assess their coding integrity.
While housing this content in a unified, publicly available environment constitutes a major step forward for the molecular modeling community, it further presents the opportunity to understand the range of validity of interatomic models and their suitability for specific target applications.
To this end, OpenKIM includes a computational pipeline that runs tests and verification checks using all available interatomic models contained within the OpenKIM Repository at \url{https://openkim.org}.
The OpenKIM Processing Pipeline is built on a set of Docker images hosted on distributed, heterogeneous hardware and utilizes open-source software to automatically run test--model and verification check--model pairs and resolve dependencies between them. The design philosophy and implementation
choices made in the development of the pipeline are discussed as well
as an example of its application to interatomic model selection.
\end{abstract}

\maketitle

\noindent \emph{The following article has been submitted to the Journal of
Chemical Physics. After it is published, it will be found at
\url{https://aip.scitation.org/journal/jcp}.}

%%%%%%%%%%%%%%%%%%%%%%%%%%%%%%%%%%%%%%%%%%%%%%%%%%%%%%%%%%%%%%%%%%%%%%%%%%%
%%%%%%%%%%%%%%%%%%%%%%%%%%%%%%%%%%%%%%%%%%%%%%%%%%%%%%%%%%%%%%%%%%%%%%%%%%%
%%%%%%%%%%%%%%%%%%%%%%%%%%%%%%%%%%%%%%%%%%%%%%%%%%%%%%%%%%%%%%%%%%%%%%%%%%%

\section{Introduction}
\label{sec:introduction}

As computational resources become more powerful, cheaper, and more prevalent, the use of molecular simulations is becoming increasingly prominent in the understanding and prediction of material properties.
The most accurate methods used in this domain are first principles approaches based on a fully quantum mechanical model of the potential energy surface, but these remain prohibitively expensive for many problems of interest.
Often, in order to reduce computational complexity, approximate interatomic models (referred to as \emph{interatomic potentials} or \emph{force fields}) are developed that eschew the electronic degrees of freedom in favor of a purely classical coarse-grained description of atomic interactions.
The predictive power of these simulations hinges delicately on a number of factors including the form of the model and its parameters, the physical properties under scrutiny, and the simulation method.

The development of new interatomic models is a daunting task requiring
a great deal of expertise and time. It is therefore common for researchers
to adopt models for their simulations developed by other groups and
published in the literature. This can be difficult in practice, since
in many cases the computer code used to generate the published results is not available
with the article and may not even by archived by the authors themselves.
Implementations of the model may exist in simulation packages, but these
are often unverified with unreliable provenance and so may not be consistent
with the published work.
This leaves other researchers to independently implement and test interatomic models based on the description found in literature, adding greatly to the barrier to adoption.

The Open Knowledgebase of Interatomic Models (OpenKIM, KIM)~\cite{tadmor:elliott:2011, tadmor:elliott:2013} established in 2009 and funded through the U.S.\ National Science Foundation aims to solve the scientific and practical issues of material simulations that use interatomic models through a comprehensive cyberinfrastructure.
OpenKIM is hosted at \url{https://openkim.org} and includes a repository for storing computer implementations of interatomic models, computational protocols to evaluate them including
tests to compute materials property predictions and verification checks to assess their coding integrity,
and first-principles and experimental results that serve as reference data
for comparison.
The computational protocols can be standalone but are typically applied through an existing molecular simulation platform (``simulator'').
The process necessary for these computations to run with an interatomic model is managed through a lightweight middleware library known as the KIM Application Programming Interface (API)~\cite{kim_api}.
The KIM API formally defines an abstract representation of the data and processing directives necessary to perform a molecular simulation, and provides a programmatic cross-language implementation capable of efficiently communicating them between models and simulators.
Any interatomic model code and simulator that conform to the KIM API standard are thus capable of functioning together seamlessly; currently supported simulators include ASAP~\cite{asap}, ASE~\cite{ase,ase2}, DL\_POLY~\cite{dlpoly}, GULP~\cite{gulp}, LAMMPS~\cite{lammps}, libatoms/QUIP~\cite{libatoms_quip}, MDStressLab++~\cite{admal:tadmor:2010,admal:tadmor:2011}, potfit~\cite{potfit:2006,potfit:2007,potfit:2015}, pyiron~\cite{pyiron}, and quasicontinuum (QC)~\cite{tadmor:ortiz:1996,tadmor:legoll:2013}.

The importance of archiving interatomic models has been recognized by others who have, in turn, established similar projects including the NIST Interatomic Potentials Repository (IPR)~\cite{nist_ipr1, nist_ipr2} and Jarvis-FF~\cite{jarvisff1,jarvisff2}.
However, there are two significant differences between these projects and OpenKIM.
First, as alluded to above, an interatomic model archived in OpenKIM is a software package that includes all the code necessary to evaluate the model to obtain the energy, forces, stresses and related values for a given atomic configuration. This should be contrasted with repositories that only archive model parameter files to be used with implementations in specific molecular simulation codes.
Archiving the model code is important, not only because it allows the model to function as a self-contained library that can be used in a portable fashion with many simulators, but also because the implementation of a model is typically complex, making it susceptible to programming errors and often requiring optimization.
This complexity gives rise to subtle effects in some cases, e.g.\ the specifics of the splines comprising the functional forms in a tabulated interatomic model have been shown to affect its predictions for some properties~\cite{wen:whalen:elliott:tadmor:2015}.
Maintaining this code (and its history) is paramount in avoiding duplicated development effort.
A second major distinction, and the focal point of this work, is that all of the models and computational protocols in OpenKIM are paired with one another and executed in a completely automated manner via a distributed, cloud-based platform known as the \emph{OpenKIM Processing Pipeline} (hereafter, ``the pipeline'').
Material property predictions computed in this fashion are inserted into a publicly accessible database alongside corresponding first-principles and experimental data, and aid in the analysis of individual models as well as the comparison of different models. These results are available through a publicly accessible mongo database hosted at \url{https://query.openkim.org} and a simplified query API through the kim-query python package~\cite{kim_query} and integrated within some simulators such as ASE~\cite{ase,ase2} and LAMMPS~\cite{lammps}.

%%%%%%%%%%%%%%%%%%%%%%%%%%%%%%%%%%%%%%%%%%%%%%%%%%%%%%%%%%%%%%%%%%%%%%%%%%%
%%%%%%%%%%%%%%%%%%%%%%%%%%%%%%%%%%%%%%%%%%%%%%%%%%%%%%%%%%%%%%%%%%%%%%%%%%%
%%%%%%%%%%%%%%%%%%%%%%%%%%%%%%%%%%%%%%%%%%%%%%%%%%%%%%%%%%%%%%%%%%%%%%%%%%%

\section{Overview}
\label{sec:overview}

\subsection{Content in KIM}
\label{sec:kimcontent}

Before turning attention to the pipeline itself, it is first necessary to survey the various types of content in OpenKIM that pass through it. (Note that below, standard KIM terminology is indicated using a san-serif font, e.g.\ \model refers to an interatomic model in the OpenKIM system.) The following are items of OpenKIM content addressed by the pipeline:
\linebreak
\begin{itemize}
  \item \model
  \item[] An algorithm representing a specific interaction between atoms, e.g.\ an interatomic potential or force field.
    There are two primary types of \models: \emph{portable models}, which can be used with any KIM API-compliant simulation code, and \emph{simulator models}, which only work with a specific simulation code.
    Portable models can either be \emph{standalone} or \emph{parameterized}.
    Standalone models consist of both a parameter file and the corresponding source code that implements the functional form of an interatomic model.
    Because the same source code is often reused across multiple parameter sets, KIM also allows it to be encapsulated in a \modeldriver, and parameterized models thus contain only parameter files and a reference to their driver.
    Simulator models also contain only a parameter file but instead of referencing a \modeldriver, they include a set of commands that invoke the implementation of a model found in a particular simulator, e.g.\ LAMMPS.
  \item \test
  \item[] A computer program that when coupled with a suitable Model, possibly including additional input, calculates a specific prediction (material property) for a particular configuration.
    Similar to the case of models, the code that performs the computation can typically be reused with different parameter sets, e.g.\ a code that calculates the lattice constant of face-centered cubic (fcc) Al could, with minor alterations, do the same for fcc Ni.
    Accordingly, a \test can either be standalone in nature or consist of a parameter file specifying the calculation that is read in by a \testdriver.
Each material property computed by a KIM \test conforms to a \emph{property definition}~\cite{kim_property_definition} schema defined by the \test developer for that property and archived in OpenKIM. This makes it possible to automatically compare property predictions across different \models and with first-principles or experimental reference data and enables dependencies between \tests (see Section~\ref{sec:dependencies}).
  \item \verificationcheck
  \item[] A computer program that when coupled with a \model examines a particular aspect of its coding correctness. This includes checks for programming errors, failures to satisfy required behaviors such as invariance principles, and determination of general characteristics of the \model's functional form such as smoothness.
    For example, a \verificationcheck might check whether the forces reported by a \model are consistent with the energy it reports, i.e.\ that the forces are the negative derivatives of the energy.
\end{itemize}
All of the above items (including \modeldrivers and \testdrivers) are assigned a unique identifier (or ``KIM ID'') in the OpenKIM repository that includes a three-digit version extension to record their evolution over time.
Further, each version is assigned its own digital object identifier (DOI) for persistent accessibility.

The objective of the pipeline is to automatically pair \tests and \verificationchecks with compatible \models and execute them.
A \test and \model are compatible and can be executed if (1) they are written for compatible versions of the KIM API, and (2) if the atomic species involved in the calculation of the \test are all supported by the \model.
\verificationchecks are designed to work with any atomic species supported by the \model, and so their compatibility is determined only based on criterion (1).
The material property instances generated by executing a specific \test--\model pair are collectively referred to as a \testresult, while the result generated by a \verificationcheck--\model pair is termed a \verificationresult.
In either case, if a pair fails to successfully generate a result, it produces an \error.
The execution time required to produce each \testresult, \verificationresult, or \error is collected and normalized with respect to a whetstone benchmark~\cite{whetstone} so as to give a hardware-independent estimation of the computing resources that were consumed.

\subsection{Pipeline Architecture}
\label{sec:architecture}

\begin{figure*}
  \includegraphics[width=0.8\textwidth]{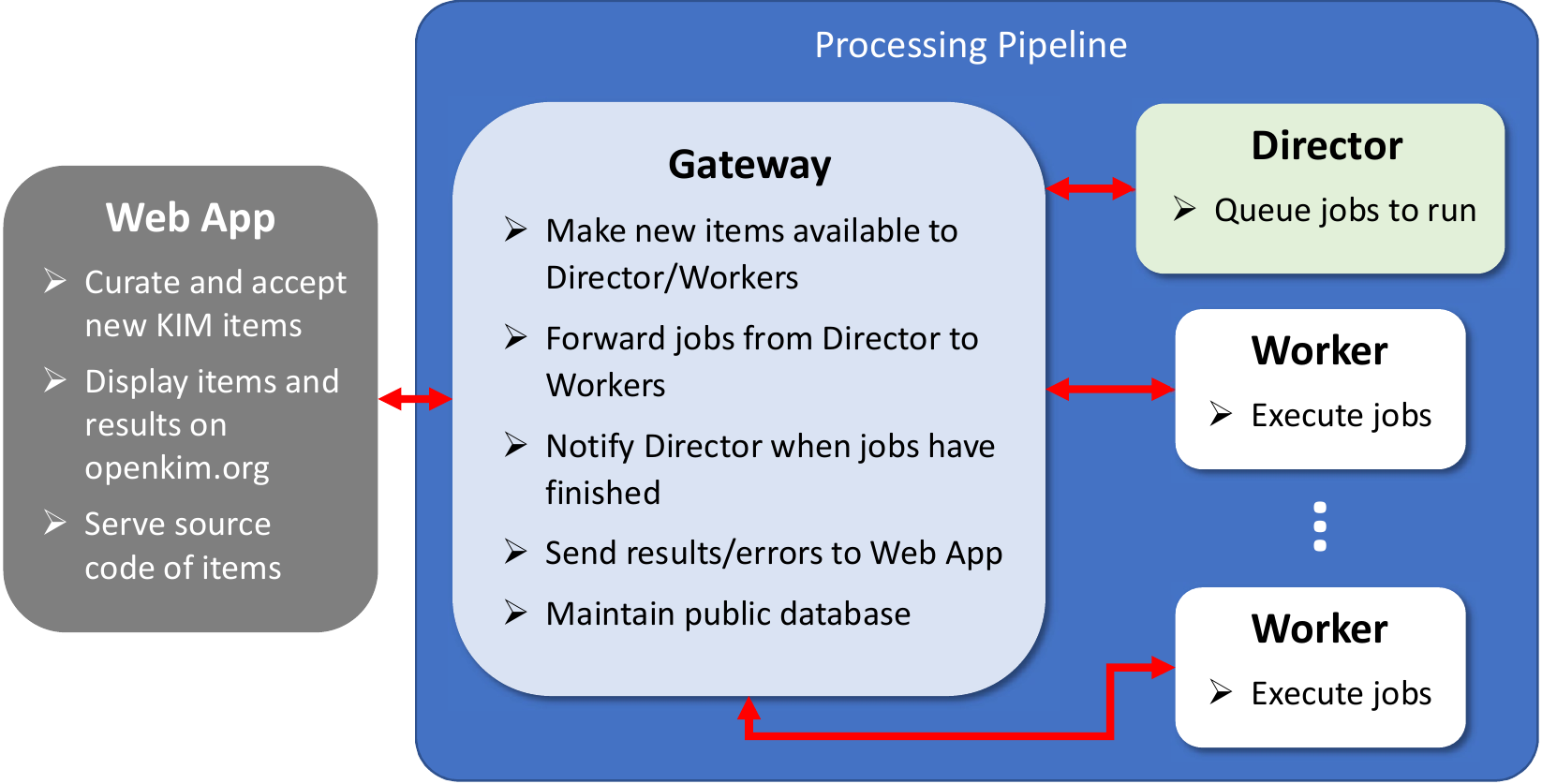}
  \caption{Abstract architecture of the pipeline and the responsibilities of each component.
           Arrows indicate connectivity.
}
  \label{fig:overview}
\end{figure*}

All \models, \tests, and \verificationchecks are submitted to the OpenKIM repository through a web application (``Web~App'') that serves the \texttt{openkim.org} domain and interfaces with the pipeline.
Once a submitted item has completed an editorial review process and been approved, a page is created for it that contains metadata associated with the item and links to its source code.
The Web~App proceeds to notify a separate \emph{Gateway} machine of the new item, which then retrieves the item and inserts it into a publicly accessible database.
Next, the Gateway sends a request to a third machine termed the \emph{Director}, whose purpose is to determine the set of all current compatible items that it can be run with.
For each compatible match that it finds, the Director creates a \emph{job} (a message corresponding to a \test--\model or \verificationcheck--\model pair that is to be run) that it communicates back to the Gateway.
Each job is claimed by one member of a fleet of \emph{Worker} machines that fetches the corresponding items from the Gateway and executes it; once a given job has completed, its results are synchronized back to the Gateway.
After inserting the results into its database, the Gateway returns them to the Web~App.
A schematic of these machines, the roles they play, and their connectivity is shown in Fig.~\ref{fig:overview}.

To make this concrete, consider a new \model for aluminum (Al) (e.g.\ an embedded-atom method (EAM) potential~\cite{daw:foiles:baskes:1993}) is added to the OpenKIM system. There are many \tests in the system designed to work with Al models. One example is a test that computes the cohesive energy (energy per atom) of Al in the face-centered cubic (fcc) structure in its equilibrium configuration. The Director will create a job coupling the Al fcc cohesive energy test with the new EAM Al potential that will be queued by the Gateway. A worker will pick up this job and perform the computation. The result will be the prediction of the new EAM potential for the cohesive energy of fcc Al. This information (encapsulated in a standard format explained below) will be returned to the Gateway and from there passed onto the Web~App for display on openkim.org. Similar calculations will be performed for all \tests that compute Al properties. In addition the new potential will be subjected to all \verificationchecks. The specifics of how such calculations are orchestrated in practice are described in Section~\ref{sec:workflow}.

Drawing on best practices in API design~\cite{reddy:2011}, the guiding principle of the pipeline architecture is encapsulation: the Web~App, Director, and Worker all have specific tasks to carry out on \models, \tests, and \verificationchecks, while the primary focus of the Gateway is to keep each of these elements isolated from one another.
This division of the pipeline into modular components based on a clear separation of responsibilities is advantageous for two reasons.
First, it reaps all of the usual benefits that accompany encapsulation.
Only simple public interfaces are exposed by each component, while private data and functions internal to them remain protected from mutation or misuse.
This enables changes of arbitrary complexity to the private data structures and functions of a component, which may be necessary for bug fixes or to accommodate changes in software dependencies, without affecting the interaction with neighboring components.
The result is comprehensible, maintainable code that is practical to adapt in response to changing design requirements.
A secondary advantage of encapsulation is that it naturally facilitates scalability.
For example, horizontal scaling of Workers or addition of Directors
to accommodate increasing computational demands is straightforward
and can be done in a dynamic fashion. High-Performance Computing (HPC)
can be accommodated by Workers geared to submission and retrieval of
tasks from HPC resources.

%%%%%%%%%%%%%%%%%%%%%%%%%%%%%%%%%%%%%%%%%%%%%%%%%%%%%%%%%%%%%%%%%%%%%%%%%%%
%%%%%%%%%%%%%%%%%%%%%%%%%%%%%%%%%%%%%%%%%%%%%%%%%%%%%%%%%%%%%%%%%%%%%%%%%%%
%%%%%%%%%%%%%%%%%%%%%%%%%%%%%%%%%%%%%%%%%%%%%%%%%%%%%%%%%%%%%%%%%%%%%%%%%%%

\section{Implementation}
\label{sec:implementation}

The implementation of the conceptual architecture described in the previous section is motivated by three main design objectives:
\begin{itemize}
    \item \textbf{Provenance} -- ability to track the origin of and recreate every \testresult, \verificationresult, and \error.
    \item \textbf{Flexibility} -- ability to run on a wide range of hardware in different physical locations and scale with computational demand.
    \item \textbf{Ease of development} -- minimization of initial and ongoing development and maintenance costs.
\end{itemize}
The first two of these objectives are satisfied with the aid of virtualization.
While this could be accomplished using full-fledged virtual machines, the pipeline is instead built upon a basis of Docker images~\cite{docker}, which have several practical advantages in the pipeline setting.
Each individual component is provisioned and stored as a version-controlled Docker image based on GNU/Linux from which a container process is spawned that runs the component.
The stack-like structure of Docker images is designed to maximize reuse of files, minimizing the amount of data that must be sent over the network when deploying new images to the components.
More importantly, because the specific Docker image used to create a component contains a complete specification of its environment, each component and any task it performs is reproducible.
In particular, the outcome of any job (\testresult or \verificationresult) can be reproduced based on the version of the Docker image used to create the Worker container that ran it.
Containerizing the pipeline components using Docker also provides fluid portability because containers can be run on nearly any modern hardware.\footnote{For HPC environments, Singularity~\cite{singularity} images can be constructed from Docker images.}
In the event that the components are run on shared hardware, the process isolation of containers minimizes the risk of interference.

The third objective in the pipeline implementation is ease of development. Because there are various operations specific to the OpenKIM framework and its contents that must be carried out, it is necessary to develop and maintain custom software for the pipeline.
The Gateway, Director, and Workers are all based on a single object-oriented codebase written in Python that features classes for the different types of KIM items (\models, \tests, etc.) as well as the Gateway, Director, and Workers themselves, that allow them to perform the tasks shown in Fig.~\ref{fig:overview}.
However, aside from this custom software, widely-used packages and protocols are used to the maximum extent possible in order to lower the burden of development and maintenance.
The rsync~\cite{rsync} utility is used to transfer the \models, \modeldrivers, \tests, \testdrivers, and \verificationchecks between the local repositories of KIM items on the Gateway, Director, and Workers.
Tornado~\cite{tornado} is used to provide an authenticated web-based HTTPS control API at \texttt{pipeline.openkim.org} accessible to the Web App for submitting new items, as well as a web interface to the public database run by the Gateway, which is implemented using MongoDB~\cite{mongo}, at \texttt{query.openkim.org}.
The Director uses SQLite~\cite{sqlite} to maintain an internal database for keeping track of jobs and dependencies between them (to be discussed in a later section).
Finally, Workers include copies of the KIM API-compliant molecular simulation codes mentioned in Section~\ref{sec:introduction}.

The most critical external software packages leveraged in the pipeline are those that connect all of its components:  Celery~\cite{celery} and RabbitMQ~\cite{rabbitmq}.
Celery is an open-source distributed task queuing framework written in Python.
In this context, a \emph{task} can be thought of as an arbitrary function to be executed on some arguments.
In the case of the pipeline, the classes that define the Gateway, Director, and Workers each have a number of member functions that perform some manner of processing on KIM items.
Those member functions that must be invoked by other components of the pipeline are thus registered as Celery tasks.
Celery prompts the actual execution of its registered tasks by way of message passing.
On each component, a Celery daemon is run that waits to receive a message requesting that it execute a specific task with some arguments.
For example, a Celery daemon runs on each Worker that waits for a message asking it to execute a specific job.
Such a message, which is created by the Director, contains as its arguments the names of the specific \test or \verificationcheck and \model that are to be run together.

Message passing in Celery is orchestrated by a \emph{message broker}.
Although multiple message brokers are available to be used with Celery, RabbitMQ was chosen for the pipeline because of its robustness and extensive feature set.
Written in Erlang, RabbitMQ implements message passing using what is known as the advanced message queuing protocol (AMQP).\footnote{Currently, RabbitMQ features native support only for AMQP version 0.9.1, employed here.}
AMQP is a protocol that adheres to the \emph{publisher--subscriber} messaging pattern.
Rather than sending messages directly from one component to another, they are placed in extensible buffers called \emph{queues} that are polled by \emph{subscribers} that acquire and process them.
In fact, messages are not even sent directly to queues, but rather to \emph{exchanges} that can implement different logic for routing messages to the queues that are bound to it.
In the pipeline, however, there is only a single exchange with three queues bound to it:  one to which the Gateway subscribes, one to which the Director subscribes, and one to which all of the Workers subscribe.
The Gateway publishes messages to the Director queue when it requests that it create jobs for a newly approved KIM item or when a job has finished running, the Director publishes the jobs it creates as messages in the Worker queue, and the Workers publish messages to the Gateway queue as they finish executing jobs.
All flow of control in the pipeline is conducted by RabbitMQ, while all execution is handled by Celery.

%%%%%%%%%%%%%%%%%%%%%%%%%%%%%%%%%%%%%%%%%%%%%%%%%%%%%%%%%%%%%%%%%%%%%%%%%%%
%%%%%%%%%%%%%%%%%%%%%%%%%%%%%%%%%%%%%%%%%%%%%%%%%%%%%%%%%%%%%%%%%%%%%%%%%%%
%%%%%%%%%%%%%%%%%%%%%%%%%%%%%%%%%%%%%%%%%%%%%%%%%%%%%%%%%%%%%%%%%%%%%%%%%%%

\section{Computational Workflow}
\label{sec:workflow}

\begin{figure*}
  \includegraphics[width=0.8\textwidth]{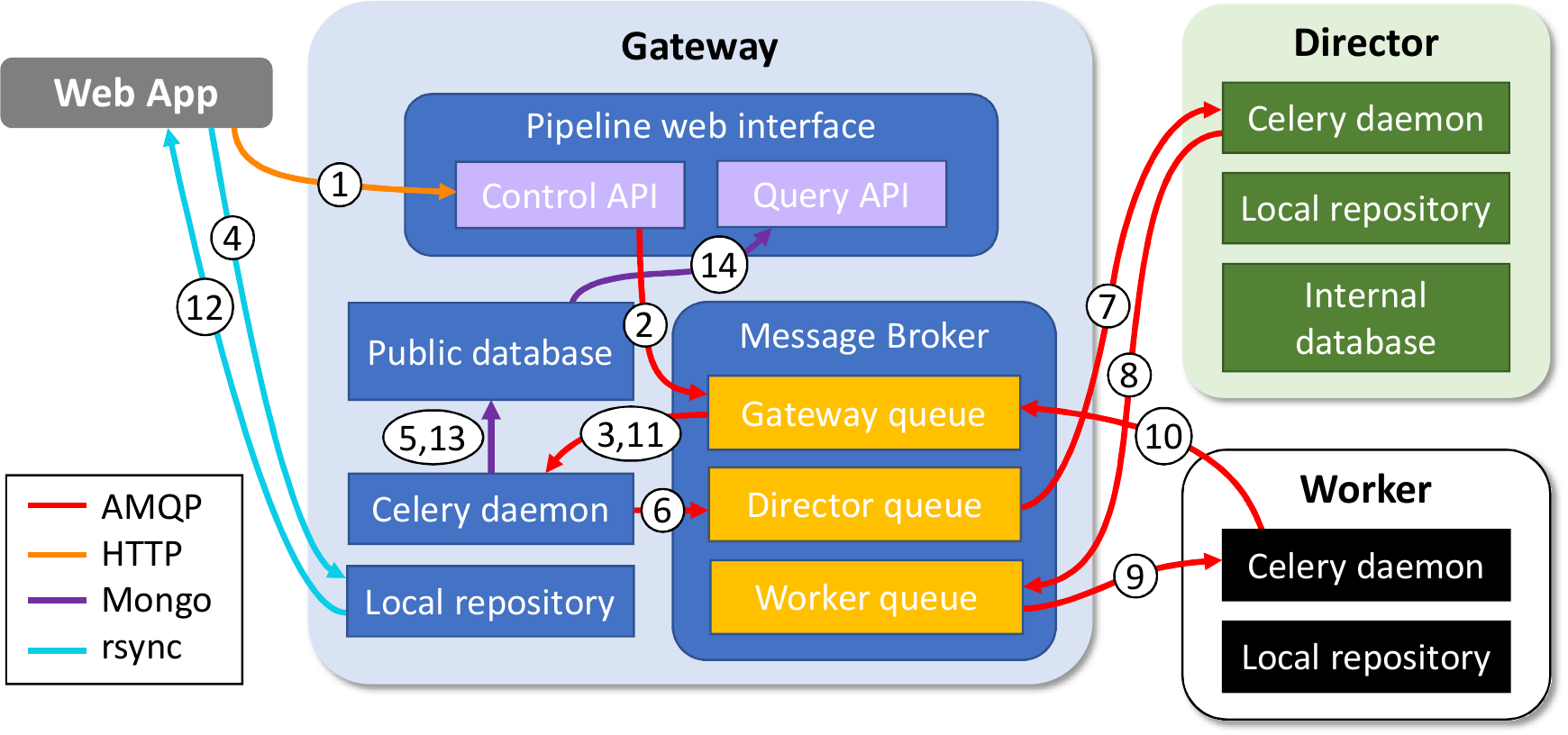}
  \caption{Internals of the pipeline components and the communication between them when a new item is submitted.
           See Section~\ref{sec:workflow} for details.
           Note that when multiple Workers are running, they all read to and write from the same queues, and the broker ensures that each job is only acquired by a single Worker.}
  \label{fig:workflow}
\end{figure*}

In order to gain a better understanding of the components of the pipeline and their internals, consider the sequence of operations that occur when a new \test is uploaded to OpenKIM and approved.
For the purposes of this example, suppose the newly approved \test, $T$, computes the lattice constant of fcc Al and there is only a single \model, $M$, for Al that exists in the OpenKIM repository.
As pictured in Fig.~\ref{fig:workflow}, the Web App begins the submission of $T$ to the pipeline by \Circled{1} notifying its Control API by sending an HTTP request to \texttt{pipeline.openkim.org}.
The pipeline control API responds by \Circled{2} placing a message on the Gateway queue indicating a new item has been submitted.
The Celery daemon running on the Gateway polls this queue and \Circled{3} acquires the message, causing it to \Circled{4} rsync the item from the official OpenKIM repository on the Web App to its own local repository.
After \Circled{5} inserting the item into the public database, the Gateway Celery daemon \Circled{6} places a message on the Director queue to inform it of the new item.
The Director Celery daemon, polling the Director queue, \Circled{7} acquires this message and rsyncs the item from the local repository of the Gateway to its own local repository.
Since the newly received item was a \test, the Director proceeds to loop over all \models that might be compatible with $T$.
Finding $M$ is compatible with $T$, the Director daemon creates a job message for the pair $T$--$M$ and \Circled{8} places it on the Worker queue.
The Worker daemon \Circled{9} acquires this message from the Worker queue and subsequently executes the job.
Once the job has finished running, the Worker announces so by \Circled{10} placing a corresponding message on the Gateway queue.
The Gateway daemon \Circled{11} acknowledges this message and rsyncs the directory containing the output of the job, which could be either a \testresult or \error, from the local repository of the Worker to its own local repository.
The Gateway daemon then \Circled{12} rsyncs the job output directory from its local repository to the Web App to be placed in the OpenKIM repository and displayed on \texttt{openkim.org}.
Finally, the Gateway daemon \Circled{13} inserts the \testresult or \error into the public-facing database where \Circled{14} it can be accessed by the Query API hosted at \texttt{query.openkim.org}.
A similar process takes place when a new \model or \verificationcheck is uploaded.

%%%%%%%%%%%%%%%%%%%%%%%%%%%%%%%%%%%%%%%%%%%%%%%%%%%%%%%%%%%%%%%%%%%%%%%%%%%
%%%%%%%%%%%%%%%%%%%%%%%%%%%%%%%%%%%%%%%%%%%%%%%%%%%%%%%%%%%%%%%%%%%%%%%%%%%
%%%%%%%%%%%%%%%%%%%%%%%%%%%%%%%%%%%%%%%%%%%%%%%%%%%%%%%%%%%%%%%%%%%%%%%%%%%

\section{Dependencies}
\label{sec:dependencies}

One subtlety not illustrated in the preceding example is that \tests in OpenKIM are allowed to make use of \testresults computed by other \tests.
Indeed, this is encouraged whenever possible because creating \tests is typically complicated and they can be expensive to run against even simple \models.
Such dependencies between \tests are made possible by the fact that all \testresults (and \verificationresults) contain, at a minimum, a file that includes one or more \emph{property instances}\footnote{Note that the kim-property python package~\cite{kim_property} can be used to create and write property instances.  A native implementation in LAMMPS is also available.}, numerical realizations of \emph{property definitions}~\cite{kim_property_definition}. Property definitions are intended to embody all physical information necessary to define a material property while ignoring any algorithmic or implementational details related to how they are computed.
Each contains a set of keys that represent physical quantities that have a well-defined data type and unit specification, and are either required to be reported in each corresponding property instance or may optionally be supplied.
For example the cohesive energy of a cubic crystal is defined by four required keys: lattice constant of the conventional unit cell, basis atom coordinates, basis atom species, and the cohesive energy itself. Optional keys include a human-readable name for the crystal type and keys for a precise Wyckoff representation of the crystal structure.
By storing \testresults in an explicit, machine-readable format in the public database of the pipeline, other \tests can use them for their own purposes with appropriately crafted queries.
These queries can be done in several ways, including simulator-native commands or the kim-query python package~\cite{kim_query}.

The existence of dependencies between \tests places restrictions on the order in which jobs can be scheduled in the pipeline.
To manage this, each \test is required to provide a file that lists which other \tests it depends on results from, which we refer to as its \emph{upstream dependencies}.\footnote{Strictly speaking, what is listed are \emph{lineages} of \tests, which encompass all versions of that \test.  The dependency is always taken to correspond to the latest existing version in that lineage.}
Conversely, the set of \tests that rely on the results of a given \test are termed its \emph{downstream dependents}.
Altogether, this means that the collection of all \tests in OpenKIM can be thought of as a directed acyclic graph.
There are two mechanisms employed by the pipeline to traverse this structure as it executes jobs, both of which are carried out by the Director: \emph{upstream resolution} and \emph{downstream resolution}.
Upstream resolution occurs when a compatible \test--\model pair is first found.
Before creating a job for the pair, the Director inspects the dependencies file of the \test.
If there are \testresults for each pairing of the \tests listed with the \model in question, the job is placed on the Worker queue.
However, if any are missing, the Director performs upstream resolution for those pairs.
This continues recursively to identify the set of all unique \test--\model pairs that are indirect upstream dependencies of the original \test--\model pair and whose own upstream dependencies are all satisfied.
Finally, jobs are created for each pair in this list and placed on the Worker queue.
Once the Gateway notifies the Director of a newly generated \testresult, downstream resolution is carried out.
The Director first reads the \test and \model used to generate the \testresult from the message placed on its queue by the Gateway.
It then searches its internal database for any \tests that are downstream dependents of the \test indicated in the \testresult message.
Any downstream dependents that have any of the others as an upstream dependency are discarded before proceeding.\footnote{This is applicable in the event where a new version of an existing \test is uploaded, which forces its downstream dependents to be rerun.  The reason is that jobs associated with the downstream dependents being removed from the list could otherwise eventually be run twice when downstream resolution is performed on the \testresults of jobs associated with the others.  However, this mechanism can fail if more complicated structures exist in the dependency graph.  A point of future work is to address this shortcoming with a global graph traversal method, e.g.\ a topological sorting algorithm, while taking care not to needlessly sequentialize jobs in independent branches.}
Each remaining downstream dependent is coupled with the \model and upstream resolution is performed on each pair in order to arrive at a unique list of \test--\model pairs to run.
Once all of the downstream dependents have been iterated over, jobs are queued for all pairs in the list.

\begin{figure*}
    \centering
    \begin{subfigure}[ht]{.25\linewidth}
      \includegraphics[width=\linewidth]{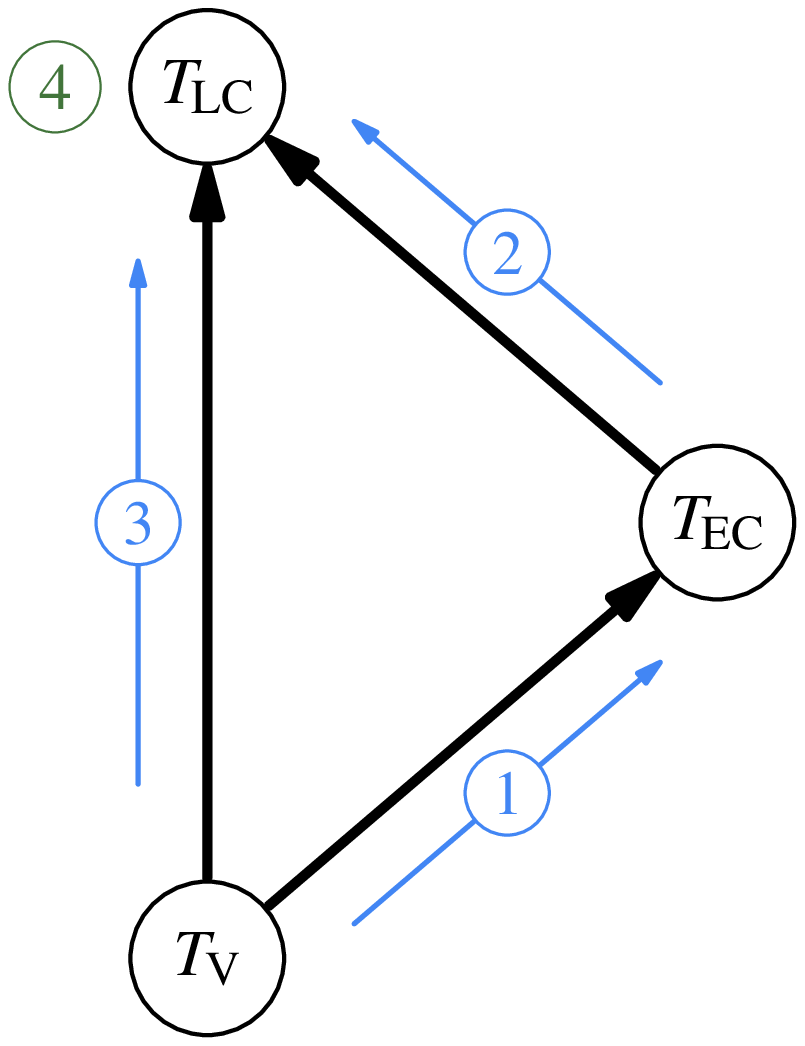}
      \caption{}
    \end{subfigure}
    \begin{subfigure}[ht]{.25\linewidth}
      \includegraphics[width=\linewidth]{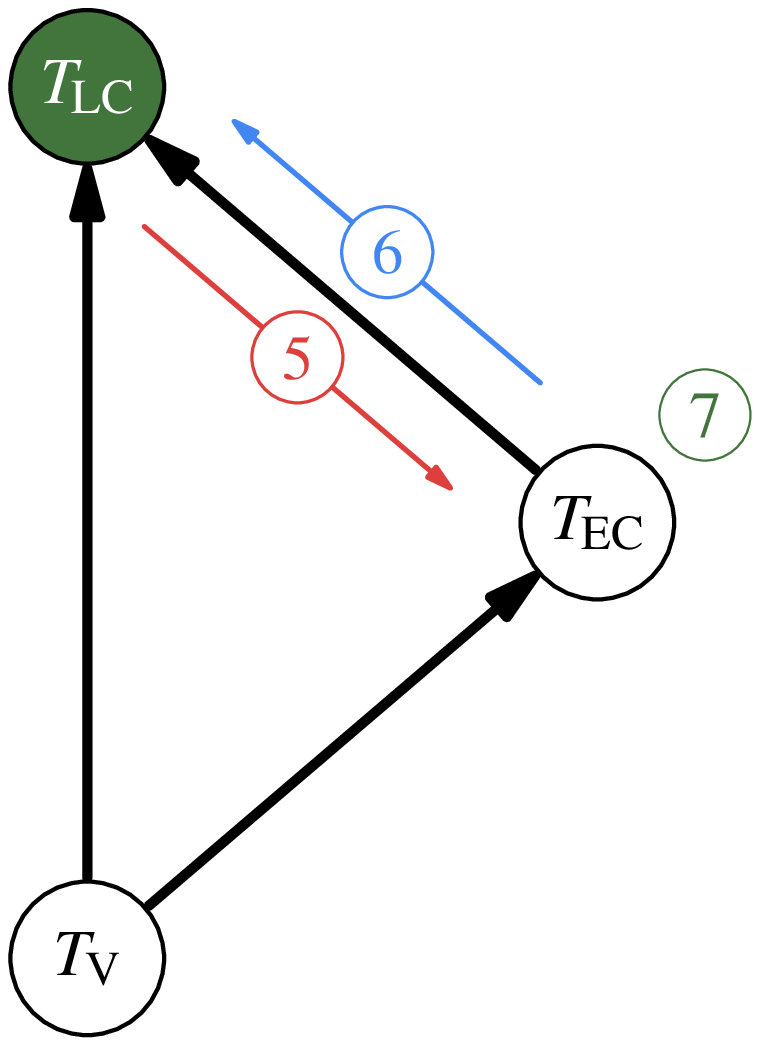}
      \caption{}
    \end{subfigure}
    \begin{subfigure}[ht]{.25\linewidth}
      \includegraphics[width=\linewidth]{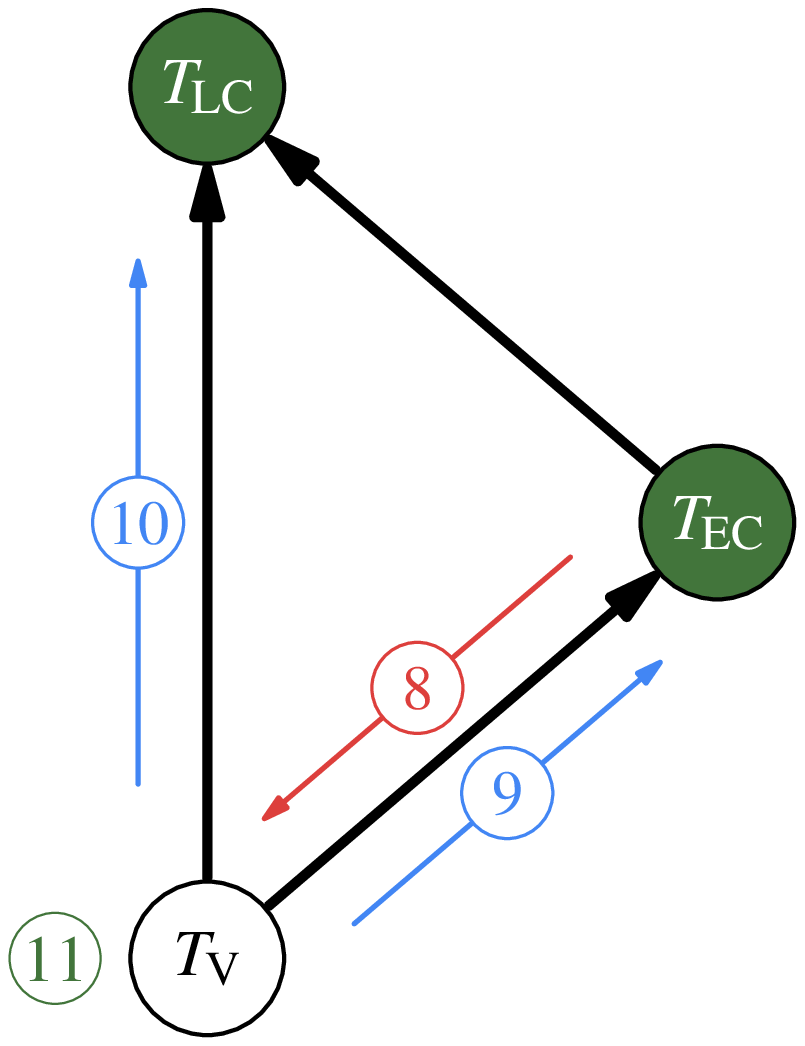}
      \caption{}
    \end{subfigure}
  \caption{Example of dependency resolution when a new \model is uploaded to the processing pipeline.
    Black arrows indicate upstream dependencies while blue and red arrows represent upstream and downstream resolution, respectively.
    (a) Upstream resolution begins from $T_{\rm V}$--$M$ and leads to $T_{\rm LC}$--$M$ being run.
    (b) Downstream resolution begins from $T_{\rm LC}$--$M$ and leads to $T_{\rm EC}$--$M$ being run.
    (c) Downstream resolution begins from $T_{\rm EC}$--$M$ and leads to $T_{\rm V}$--$M$ being run.
}
  \label{fig:dependencies}
\end{figure*}

An explicit example is shown in Fig.~\ref{fig:dependencies}.
Suppose there exist several \tests that calculate properties of fcc Al at zero temperature: one that computes the lattice constant ($T_{\rm LC}$), one that computes the elastic constants ($T_{\rm EC}$), and one that computes the stress field surrounding a monovacancy using a linear elastic model (${T_{\rm V}}$).
The elastic constants \test has the lattice constant \test as its upstream dependency, whereas the vacancy \test has both the elastic constants and lattice constants \tests as its upstream dependencies.
Next, assume that a new \model $M$ for Al has been uploaded to the OpenKIM Repository.
When the Director is notified of the new model, it begins looping over all current \tests to determine which of them are compatible with the model.
For the purposes of this example, assume that the first \test the Director visits is $T_{\rm V}$.
The first phase of dependency resolution is shown in Fig.~\ref{fig:dependencies}(a) (the circled numbers below refer to dependency resolution steps in the figure).
After determining it is a compatible match with $M$, the Director begins iterating over its upstream dependencies to see if they are satisfied.
In the case of a \test with multiple dependencies, the order in which it lists them in its dependencies file is arbitrary.
Supposing that $T_{\rm EC}$ is listed first, the Director attempts to match it with $M$ and perform upstream resolution on this pair \Circled{1}.
Although it is found to be compatible, \Circled{2} the Director finds that the upstream dependency of $T_{\rm EC}$, $T_{\rm LC}$, has not yet been run against $M$.
Recursing once more, the Director matches $T_{\rm LC}$ with the $M$ and performs upstream resolution on the pair.
This time, since $T_{\rm LC}$ has no upstream dependencies, it is determined that the pair is ready to run and it is passed back down to the original upstream resolution that was started at $T_{\rm V}$ to be added to the run list.
Having looped over $T_{\rm EC}$ during the original upstream resolution, \Circled{3} the Director attempts upstream resolution on $T_{\rm V}$'s other dependency $T_{\rm LC}$.
Although it finds that $T_{\rm LC}$ is ready to run against $M$, the pair is already found in the run list so it is ignored.
Having completed the upstream resolution from $T_{\rm V}$, \Circled{4} a job is created for the pair $T_{\rm LC}$--$M$ and pushed to the Worker queue.
The next phase of dependency resolution is shown in Fig.~\ref{fig:dependencies}(b).
Assuming the job produces a \testresult (rather than an \error), \Circled{5} the Director is notified and begins downstream resolution for $T_{\rm LC}$.
Observing that $T_{\rm EC}$ is an upstream dependency of $T_{\rm V}$, the latter is discarded from consideration, leaving only downstream resolution to $T_{\rm EC}$.
\Circled{6} Upstream resolution on the pair $T_{\rm EC}$--$M$ confirms that $T_{\rm LC}$ has been run and that there are no other upstream dependencies, so \Circled{7} a job for the pair is created and queued.
The final phase of dependency resolution is shown in Fig.~\ref{fig:dependencies}(c).
Once the \testresult corresponding to $T_{\rm EC}$--$M$ is returned to the Director, \Circled{8} downstream resolution leads the Director to $T_{\rm EC}$'s one downstream dependent $T_{\rm V}$.
Now, \Circled{9}--\Circled{10} upstream resolution of $T_{\rm V}$--$M$ indicates that all of its upstream dependencies are met and \Circled{11} it is run.

%%%%%%%%%%%%%%%%%%%%%%%%%%%%%%%%%%%%%%%%%%%%%%%%%%%%%%%%%%%%%%%%%%%%%%%%%%%
%%%%%%%%%%%%%%%%%%%%%%%%%%%%%%%%%%%%%%%%%%%%%%%%%%%%%%%%%%%%%%%%%%%%%%%%%%%
%%%%%%%%%%%%%%%%%%%%%%%%%%%%%%%%%%%%%%%%%%%%%%%%%%%%%%%%%%%%%%%%%%%%%%%%%%%

\section{Application to Model Selection}
\label{sec:application}

A practical application of the data produced by the OpenKIM pipeline is the selection of an interatomic model for a specific target application.
To aid in this process, the ``KIM Compare'' tool~\cite{kim_compare} aggregates \testresults for a set of properties of interest for a range of \models and displays them to the user in the form of dynamic tables and graphs.
The first step is to identify a set of $N_\mathrm{props}$ properties deemed important for a model to reproduce accurately for the fidelity of the target application, and for which first principles or experimental reference data is available. The absolute relative error between the model prediction and the reference data for each property is defined as
\begin{equation}
  e^M_p := \left| \frac{V^M_p - R_p}{R_p} \right|,
\end{equation}
where $V^M_p$ is the prediction of model $M$ for property $p$ and $R_p$ is a reference value. In order to compare between models, a cost function is defined as a weighted sum (with weights $w_p > 0$) over the relative errors, so that for model $M$ the error cost is
\begin{equation}
  \zeta^M := \frac{ \sum_{p=1}^{N_\mathrm{props}} w_p  e^M_p }{ \sum_{p=1}^{N_\mathrm{props}} w_p }.
  \label{eq:zeta}
\end{equation}
The lower the cost $\zeta^M$ the more accurate the model is overall. The weights in Eq.~(\ref{eq:zeta}) are selected based on domain expertise and intuition regarding the relative importance of the properties for the target application. An area of active research in the OpenKIM project is to develop more rigorous methods for identifying properties of importance and associated weights for an arbitrary target application~\cite{karls:2016}.

In addition to accuracy, computational cost is also an important consideration when selecting a model. As a measure of the speed of a model, its average execution time over all $N_\mathrm{props}$ properties is computed. For model $M$ this is
\begin{equation}
  \bar{t}^M := \frac{1}{N_\mathrm{props}} \sum_{p=1}^{N_\mathrm{props}} t^M_p,
  \label{eq:tave}
\end{equation}
where $t^M_p$ is the execution time for computing property $p$ with model $M$ normalized by the whetstone benchmark (see Section~\ref{sec:overview}). By using normalized time, computations performed on Workers running on different architectures are considered on equal footing.

A model can be selected from a pool of available candidates by examining the results from Eqns.~(\ref{eq:zeta}) and~(\ref{eq:tave}) on a cost versus time plot generated by the KIM~Compare tool. A recent real-world example of usage of this tool was the selection of a copper (Cu) model for a large-scale molecular dynamics simulation of crystal plasticity at Lawrence Livermore National Laboratory (LANL)~\cite{bulatov_privcomm, zepeda:2019, zepeda:2017}.
The objective was to find a model that was as inexpensive as possible in order to maximize the size of the simulation while still being sufficiently accurate for the material properties being studied.
Crystal plasticity in fcc crystals is governed by dislocation nucleation and interaction.  Key properties for obtaining correct behavior include the elastic constants that govern the long-range interaction between dislocations, the intrinsic stacking fault energy that governs the splitting distance in dissociated dislocation cores, and basic crystal properties including the equilibrium lattice constant and cohesive energy.  In addition, it is important that the likelihood of dislocation nucleation relative to competing mechanisms such as deformation twinning or brittle fracture is captured. This is governed by the unstable stacking energy~\cite{rice:1992}, unstable twinning energy~\cite{tadmor:hai:2003}, and surface energies of potential cleavage planes.

\begin{figure*}
  \includegraphics[width=0.8\linewidth]{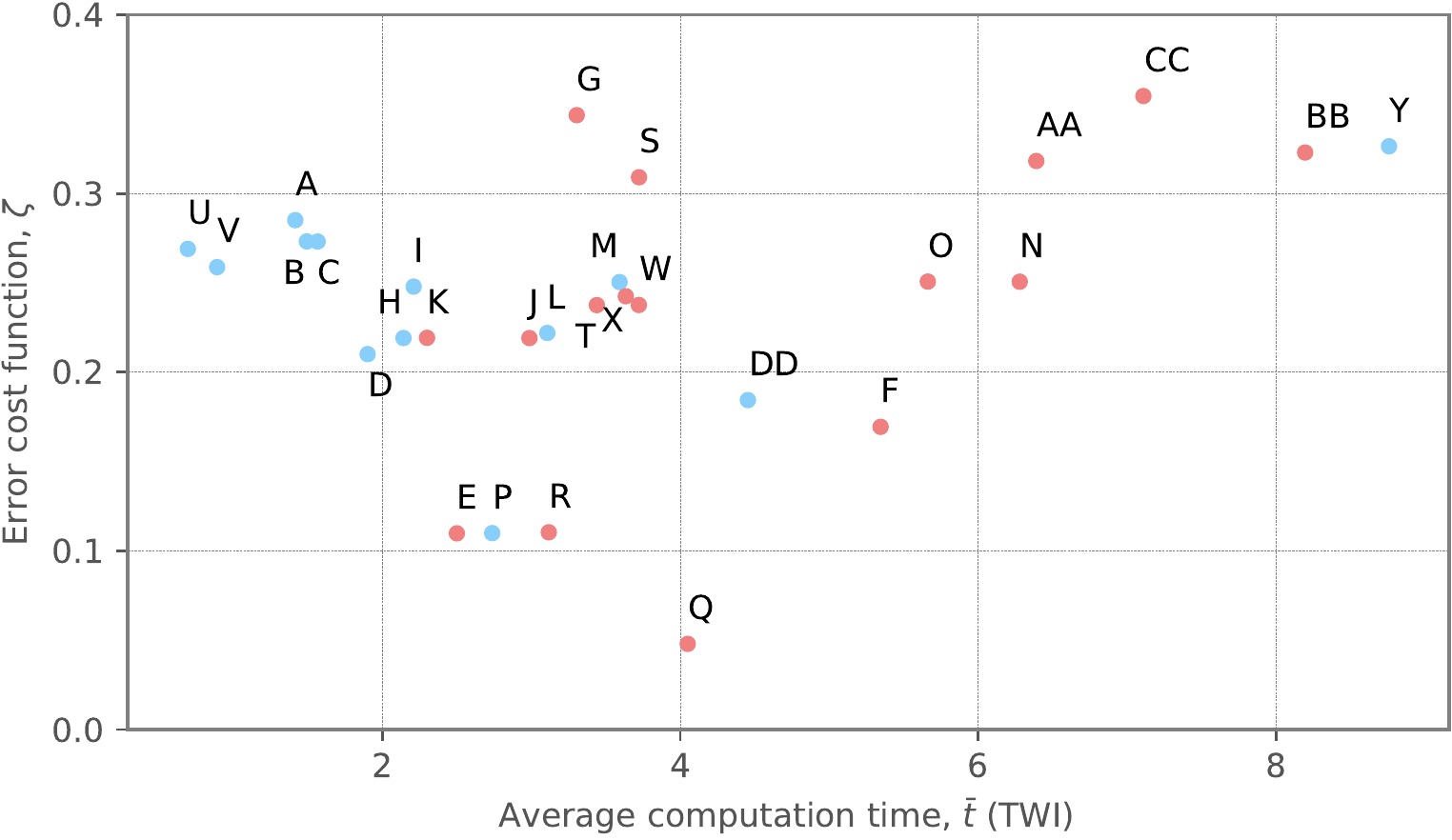}
  \caption{Cost function $\zeta^M$ defined in Eq.~(\ref{eq:zeta}) versus the average computation time in Eq.~(\ref{eq:tave}), given in units of Tera-Whetstone Instructions (TWI), for 30 EAM potentials for Cu archived in OpenKIM and 13 material properties of fcc Cu.  Monospecies models (blue) display similar accuracy to multispecies models (red), but are typically computationally less expensive. See the Supplementary Material for a definition of the model labels.
}
  \label{fig:model_comparison}
\end{figure*}

The cost versus computation time for 30 EAM potentials archived in OpenKIM are shown in Fig.~\ref{fig:model_comparison}. (See the Supplementary Material for a spreadsheet containing the full listing of the material properties, models, and reference data used. Weights can be manipulated in the spreadsheet to see how this affects model selection.) Only EAM potentials were considered, since they are known to provide acceptable accuracy for fcc metals and are significantly less expensive than more accurate options. The differences in computation time between the EAM models is related to details such as the employed cutoff radius, EAM functional forms, and in the case of tabulated functions, the number of data points. Based on these results, model ``P'' by Mishin \emph{et al.}~\cite{MO_346334655118_005, MD_120291908751_005, MO_346334655118_005a} was selected by the LANL researchers because it provided a good compromise in terms of relative speed and accuracy.

%%%%%%%%%%%%%%%%%%%%%%%%%%%%%%%%%%%%%%%%%%%%%%%%%%%%%%%%%%%%%%%%%%%%%%%%%%%
%%%%%%%%%%%%%%%%%%%%%%%%%%%%%%%%%%%%%%%%%%%%%%%%%%%%%%%%%%%%%%%%%%%%%%%%%%%
%%%%%%%%%%%%%%%%%%%%%%%%%%%%%%%%%%%%%%%%%%%%%%%%%%%%%%%%%%%%%%%%%%%%%%%%%%%

\section{Conclusions and Future Work}

The OpenKIM Pipeline is a distributed infrastructure to orchestrate the computation of compatible \models, \tests, and \verificationchecks in the OpenKIM repository.
This infrastructure is divided into different encapsulated components based on a clear separation of responsibilities.
Each component is implemented as a Docker container, providing reproducibility of their environment and the tasks they perform, as well as portability across heterogeneous hardware.
Moreover, common software packages and protocols are leveraged not only in the majority of the individual components but also in the networking that allows them to communicate with one another.
Altogether, the design choices made support the project-wide goals of provenance, flexibility, and ease of development. The results from the calculations performed by the pipeline are archived at \url{openkim.org} and are used by the KIM~Compare tool to help users select models for applications of interest.

Further work is needed to implement a more sophisticated algorithm for job scheduling that excludes the possibility of jobs being rerun unnecessarily in the case of pathological dependencies structures.
Support must also be added for jobs that require HPC resources, including those external to the pipeline itself.
This may entail a revision of the containerization approach so that a Docker image is created for each individual job~\cite{containerintegration}.
It also brings forward the need for a job prioritization system, which might take into account profiling information for jobs previously run for each \test in order to predict the computational demand of future jobs.
Vertical scaling of the individual components of the pipeline is becoming increasingly important to accommodate increased community uptake.
In addition, growth in the size of the OpenKIM repository highlights the need for automated horizontal scaling based on work load. Finally, the development of intelligent tools for model comparison and selection that can assist users in this process remains a challenging and important area for continuing work.

%%%%%%%%%%%%%%%%%%%%%%%%%%%%%%%%%%%%%%%%%%%%%%%%%%%%%%%%%%%%%%%%%%%%%%%%%%%
%%%%%%%%%%%%%%%%%%%%%%%%%%%%%%%%%%%%%%%%%%%%%%%%%%%%%%%%%%%%%%%%%%%%%%%%%%%
%%%%%%%%%%%%%%%%%%%%%%%%%%%%%%%%%%%%%%%%%%%%%%%%%%%%%%%%%%%%%%%%%%%%%%%%%%%

\begin{acknowledgments}
This research was partly supported by the National Science Foundation
(NSF) under grants Nos.~DMR-1834251 and DMR-1834332.
The authors acknowledge the Minnesota Supercomputing Institute (MSI) at
the University of Minnesota for providing resources that contributed to the
results reported in this paper. The authors thank Ronald Miller, Noam Bernstein,
Mingjian Wen and Yaser Afshar for helpful discussions and for contributing
to this effort.
\end{acknowledgments}

\nocite{*}
\bibliography{kim-pipeline}

%merlin.mbs aipnum4-1.bst 2010-07-25 4.21a (PWD, AO, DPC) hacked
%Control: key (0)
%Control: author (8) initials jnrlst
%Control: editor formatted (1) identically to author
%Control: production of article title (-1) disabled
%Control: page (0) single
%Control: year (1) truncated
%Control: production of eprint (0) enabled
\begin{thebibliography}{55}%
\makeatletter
\providecommand \@ifxundefined [1]{%
 \@ifx{#1\undefined}
}%
\providecommand \@ifnum [1]{%
 \ifnum #1\expandafter \@firstoftwo
 \else \expandafter \@secondoftwo
 \fi
}%
\providecommand \@ifx [1]{%
 \ifx #1\expandafter \@firstoftwo
 \else \expandafter \@secondoftwo
 \fi
}%
\providecommand \natexlab [1]{#1}%
\providecommand \enquote  [1]{``#1''}%
\providecommand \bibnamefont  [1]{#1}%
\providecommand \bibfnamefont [1]{#1}%
\providecommand \citenamefont [1]{#1}%
\providecommand \href@noop [0]{\@secondoftwo}%
\providecommand \href [0]{\begingroup \@sanitize@url \@href}%
\providecommand \@href[1]{\@@startlink{#1}\@@href}%
\providecommand \@@href[1]{\endgroup#1\@@endlink}%
\providecommand \@sanitize@url [0]{\catcode `\\12\catcode `\$12\catcode
  `\&12\catcode `\#12\catcode `\^12\catcode `\_12\catcode `\%12\relax}%
\providecommand \@@startlink[1]{}%
\providecommand \@@endlink[0]{}%
\providecommand \url  [0]{\begingroup\@sanitize@url \@url }%
\providecommand \@url [1]{\endgroup\@href {#1}{\urlprefix }}%
\providecommand \urlprefix  [0]{URL }%
\providecommand \Eprint [0]{\href }%
\providecommand \doibase [0]{http://dx.doi.org/}%
\providecommand \selectlanguage [0]{\@gobble}%
\providecommand \bibinfo  [0]{\@secondoftwo}%
\providecommand \bibfield  [0]{\@secondoftwo}%
\providecommand \translation [1]{[#1]}%
\providecommand \BibitemOpen [0]{}%
\providecommand \bibitemStop [0]{}%
\providecommand \bibitemNoStop [0]{.\EOS\space}%
\providecommand \EOS [0]{\spacefactor3000\relax}%
\providecommand \BibitemShut  [1]{\csname bibitem#1\endcsname}%
\let\auto@bib@innerbib\@empty
%</preamble>
\bibitem [{\citenamefont {Tadmor}\ \emph {et~al.}(2011)\citenamefont {Tadmor},
  \citenamefont {Elliott}, \citenamefont {Sethna}, \citenamefont {Miller},\
  and\ \citenamefont {Becker}}]{tadmor:elliott:2011}%
  \BibitemOpen
  \bibfield  {author} {\bibinfo {author} {\bibfnamefont {E.~B.}\ \bibnamefont
  {Tadmor}}, \bibinfo {author} {\bibfnamefont {R.~S.}\ \bibnamefont {Elliott}},
  \bibinfo {author} {\bibfnamefont {J.~P.}\ \bibnamefont {Sethna}}, \bibinfo
  {author} {\bibfnamefont {R.~E.}\ \bibnamefont {Miller}}, \ and\ \bibinfo
  {author} {\bibfnamefont {C.~A.}\ \bibnamefont {Becker}},\ }\href {\doibase
  10.1007/s11837-011-0102-6} {\bibfield  {journal} {\bibinfo  {journal} {JOM}\
  }\textbf {\bibinfo {volume} {63}},\ \bibinfo {pages} {17} (\bibinfo {year}
  {2011})}\BibitemShut {NoStop}%
\bibitem [{\citenamefont {Tadmor}\ \emph
  {et~al.}(2013{\natexlab{a}})\citenamefont {Tadmor}, \citenamefont {Elliott},
  \citenamefont {Phillpot},\ and\ \citenamefont
  {Sinnott}}]{tadmor:elliott:2013}%
  \BibitemOpen
  \bibfield  {author} {\bibinfo {author} {\bibfnamefont {E.~B.}\ \bibnamefont
  {Tadmor}}, \bibinfo {author} {\bibfnamefont {R.~S.}\ \bibnamefont {Elliott}},
  \bibinfo {author} {\bibfnamefont {S.~R.}\ \bibnamefont {Phillpot}}, \ and\
  \bibinfo {author} {\bibfnamefont {S.~B.}\ \bibnamefont {Sinnott}},\ }\href
  {\doibase 10.1016/j.cossms.2013.10.004} {\bibfield  {journal} {\bibinfo
  {journal} {COSSMS}\ }\textbf {\bibinfo {volume} {17}},\ \bibinfo {pages}
  {298} (\bibinfo {year} {2013}{\natexlab{a}})}\BibitemShut {NoStop}%
\bibitem [{\citenamefont {Elliott}\ and\ \citenamefont
  {Tadmor}(2011)}]{kim_api}%
  \BibitemOpen
  \bibfield  {author} {\bibinfo {author} {\bibfnamefont {R.~S.}\ \bibnamefont
  {Elliott}}\ and\ \bibinfo {author} {\bibfnamefont {E.~B.}\ \bibnamefont
  {Tadmor}},\ }\href {\doibase 10.25950/ff8f563a} {\enquote {\bibinfo {title}
  {{Knowledgebase of Interatomic Models (KIM) Application Programming Interface
  (API)}},}\ } (\bibinfo {year} {2011})\BibitemShut {NoStop}%
\bibitem [{asa()}]{asap}%
  \BibitemOpen
  \href@noop {} {\enquote {\bibinfo {title} {{A}sap - {A}s {S}oon {A}s
  {P}ossible},}\ }\bibinfo {howpublished}
  {\url{https://wiki.fysik.dtu.dk/asap}}\BibitemShut {NoStop}%
\bibitem [{\citenamefont {Larsen}\ \emph {et~al.}(2017)\citenamefont {Larsen},
  \citenamefont {Mortensen}, \citenamefont {Blomqvist}, \citenamefont
  {Castelli}, \citenamefont {Christensen}, \citenamefont {Du{\l}ak},
  \citenamefont {Friis}, \citenamefont {Groves}, \citenamefont {Hammer},
  \citenamefont {Hargus}, \citenamefont {Hermes}, \citenamefont {Jennings},
  \citenamefont {Jensen}, \citenamefont {Kermode}, \citenamefont {Kitchin},
  \citenamefont {Kolsbjerg}, \citenamefont {Kubal}, \citenamefont {Kaasbjerg},
  \citenamefont {Lysgaard}, \citenamefont {Maronsson}, \citenamefont {Maxson},
  \citenamefont {Olsen}, \citenamefont {Pastewka}, \citenamefont {Peterson},
  \citenamefont {Rostgaard}, \citenamefont {Schi{\o}tz}, \citenamefont
  {Sch{\"{u}}tt}, \citenamefont {Strange}, \citenamefont {Thygesen},
  \citenamefont {Vegge}, \citenamefont {Vilhelmsen}, \citenamefont {Walter},
  \citenamefont {Zeng},\ and\ \citenamefont {Jacobsen}}]{ase}%
  \BibitemOpen
  \bibfield  {author} {\bibinfo {author} {\bibfnamefont {A.~H.}\ \bibnamefont
  {Larsen}}, \bibinfo {author} {\bibfnamefont {J.~J.}\ \bibnamefont
  {Mortensen}}, \bibinfo {author} {\bibfnamefont {J.}~\bibnamefont
  {Blomqvist}}, \bibinfo {author} {\bibfnamefont {I.~E.}\ \bibnamefont
  {Castelli}}, \bibinfo {author} {\bibfnamefont {R.}~\bibnamefont
  {Christensen}}, \bibinfo {author} {\bibfnamefont {M.}~\bibnamefont
  {Du{\l}ak}}, \bibinfo {author} {\bibfnamefont {J.}~\bibnamefont {Friis}},
  \bibinfo {author} {\bibfnamefont {M.~N.}\ \bibnamefont {Groves}}, \bibinfo
  {author} {\bibfnamefont {B.}~\bibnamefont {Hammer}}, \bibinfo {author}
  {\bibfnamefont {C.}~\bibnamefont {Hargus}}, \bibinfo {author} {\bibfnamefont
  {E.~D.}\ \bibnamefont {Hermes}}, \bibinfo {author} {\bibfnamefont {P.~C.}\
  \bibnamefont {Jennings}}, \bibinfo {author} {\bibfnamefont {P.~B.}\
  \bibnamefont {Jensen}}, \bibinfo {author} {\bibfnamefont {J.}~\bibnamefont
  {Kermode}}, \bibinfo {author} {\bibfnamefont {J.~R.}\ \bibnamefont
  {Kitchin}}, \bibinfo {author} {\bibfnamefont {E.~L.}\ \bibnamefont
  {Kolsbjerg}}, \bibinfo {author} {\bibfnamefont {J.}~\bibnamefont {Kubal}},
  \bibinfo {author} {\bibfnamefont {K.}~\bibnamefont {Kaasbjerg}}, \bibinfo
  {author} {\bibfnamefont {S.}~\bibnamefont {Lysgaard}}, \bibinfo {author}
  {\bibfnamefont {J.~B.}\ \bibnamefont {Maronsson}}, \bibinfo {author}
  {\bibfnamefont {T.}~\bibnamefont {Maxson}}, \bibinfo {author} {\bibfnamefont
  {T.}~\bibnamefont {Olsen}}, \bibinfo {author} {\bibfnamefont
  {L.}~\bibnamefont {Pastewka}}, \bibinfo {author} {\bibfnamefont
  {A.}~\bibnamefont {Peterson}}, \bibinfo {author} {\bibfnamefont
  {C.}~\bibnamefont {Rostgaard}}, \bibinfo {author} {\bibfnamefont
  {J.}~\bibnamefont {Schi{\o}tz}}, \bibinfo {author} {\bibfnamefont
  {O.}~\bibnamefont {Sch{\"{u}}tt}}, \bibinfo {author} {\bibfnamefont
  {M.}~\bibnamefont {Strange}}, \bibinfo {author} {\bibfnamefont {K.~S.}\
  \bibnamefont {Thygesen}}, \bibinfo {author} {\bibfnamefont {T.}~\bibnamefont
  {Vegge}}, \bibinfo {author} {\bibfnamefont {L.}~\bibnamefont {Vilhelmsen}},
  \bibinfo {author} {\bibfnamefont {M.}~\bibnamefont {Walter}}, \bibinfo
  {author} {\bibfnamefont {Z.}~\bibnamefont {Zeng}}, \ and\ \bibinfo {author}
  {\bibfnamefont {K.~W.}\ \bibnamefont {Jacobsen}},\ }\href {\doibase
  10.1088/1361-648x/aa680e} {\bibfield  {journal} {\bibinfo  {journal} {Journal
  of Physics: Condensed Matter}\ }\textbf {\bibinfo {volume} {29}},\ \bibinfo
  {pages} {273002} (\bibinfo {year} {2017})}\BibitemShut {NoStop}%
\bibitem [{\citenamefont {{Bahn, S.R. and Jacobsen, Karsten W.}}(2002)}]{ase2}%
  \BibitemOpen
  \bibfield  {author} {\bibinfo {author} {\bibnamefont {{Bahn, S.R. and
  Jacobsen, Karsten W.}}},\ }\href {\doibase 10.1109/5992.998641} {\bibfield
  {journal} {\bibinfo  {journal} {Computing in Science Engineering}\ }\textbf
  {\bibinfo {volume} {4}},\ \bibinfo {pages} {56} (\bibinfo {year}
  {2002})}\BibitemShut {NoStop}%
\bibitem [{\citenamefont {Todorov}\ \emph {et~al.}(2006)\citenamefont
  {Todorov}, \citenamefont {Smith}, \citenamefont {Trachenko},\ and\
  \citenamefont {Dove}}]{dlpoly}%
  \BibitemOpen
  \bibfield  {author} {\bibinfo {author} {\bibfnamefont {I.~T.}\ \bibnamefont
  {Todorov}}, \bibinfo {author} {\bibfnamefont {W.}~\bibnamefont {Smith}},
  \bibinfo {author} {\bibfnamefont {K.}~\bibnamefont {Trachenko}}, \ and\
  \bibinfo {author} {\bibfnamefont {M.~T.}\ \bibnamefont {Dove}},\ }\href
  {\doibase 10.1039/B517931A} {\bibfield  {journal} {\bibinfo  {journal}
  {Journal of Materials Chemistry}\ }\textbf {\bibinfo {volume} {16}},\
  \bibinfo {pages} {1911} (\bibinfo {year} {2006})}\BibitemShut {NoStop}%
\bibitem [{\citenamefont {Gale}(1997)}]{gulp}%
  \BibitemOpen
  \bibfield  {author} {\bibinfo {author} {\bibfnamefont {J.~D.}\ \bibnamefont
  {Gale}},\ }\href {\doibase 10.1039/A606455H} {\bibfield  {journal} {\bibinfo
  {journal} {Journal of the Chemical Society, Faraday Transactions}\ }\textbf
  {\bibinfo {volume} {93}},\ \bibinfo {pages} {629} (\bibinfo {year}
  {1997})}\BibitemShut {NoStop}%
\bibitem [{\citenamefont {Plimpton}(1995)}]{lammps}%
  \BibitemOpen
  \bibfield  {author} {\bibinfo {author} {\bibfnamefont {S.}~\bibnamefont
  {Plimpton}},\ }\href {\doibase 10.1006/jcph.1995.1039} {\bibfield  {journal}
  {\bibinfo  {journal} {Journal of Computational Physics}\ }\textbf {\bibinfo
  {volume} {117}},\ \bibinfo {pages} {1 } (\bibinfo {year} {1995})}\BibitemShut
  {NoStop}%
\bibitem [{\citenamefont {Bart\'{o}k}\ \emph {et~al.}()\citenamefont
  {Bart\'{o}k} \emph {et~al.}}]{libatoms_quip}%
  \BibitemOpen
  \bibfield  {author} {\bibinfo {author} {\bibfnamefont {A.~P.}\ \bibnamefont
  {Bart\'{o}k}} \emph {et~al.},\ }\href@noop {} {\enquote {\bibinfo {title}
  {{libAtoms+QUIP: A software library for carrying out molecular dynamics
  simulations}},}\ }\bibinfo {howpublished}
  {\url{http://www.libatoms.org/}}\BibitemShut {NoStop}%
\bibitem [{\citenamefont {Admal}\ and\ \citenamefont
  {Tadmor}(2010)}]{admal:tadmor:2010}%
  \BibitemOpen
  \bibfield  {author} {\bibinfo {author} {\bibfnamefont {N.~C.}\ \bibnamefont
  {Admal}}\ and\ \bibinfo {author} {\bibfnamefont {E.~B.}\ \bibnamefont
  {Tadmor}},\ }\href {\doibase 10.1007/s10659-010-9249-6} {\bibfield  {journal}
  {\bibinfo  {journal} {Journal of Elasticity}\ }\textbf {\bibinfo {volume}
  {100}},\ \bibinfo {pages} {63} (\bibinfo {year} {2010})}\BibitemShut
  {NoStop}%
\bibitem [{\citenamefont {Admal}\ and\ \citenamefont
  {Tadmor}(2011)}]{admal:tadmor:2011}%
  \BibitemOpen
  \bibfield  {author} {\bibinfo {author} {\bibfnamefont {N.~C.}\ \bibnamefont
  {Admal}}\ and\ \bibinfo {author} {\bibfnamefont {E.~B.}\ \bibnamefont
  {Tadmor}},\ }\href {\doibase 10.1063/1.3582905} {\bibfield  {journal}
  {\bibinfo  {journal} {Journal of Chemical Physics}\ }\textbf {\bibinfo
  {volume} {134}},\ \bibinfo {pages} {184106} (\bibinfo {year}
  {2011})}\BibitemShut {NoStop}%
\bibitem [{\citenamefont {Brommer}\ and\ \citenamefont
  {G\"{a}hler}(2006)}]{potfit:2006}%
  \BibitemOpen
  \bibfield  {author} {\bibinfo {author} {\bibfnamefont {P.}~\bibnamefont
  {Brommer}}\ and\ \bibinfo {author} {\bibfnamefont {F.}~\bibnamefont
  {G\"{a}hler}},\ }\href {\doibase 10.1080/14786430500333349} {\bibfield
  {journal} {\bibinfo  {journal} {Philosophical Magazine}\ }\textbf {\bibinfo
  {volume} {86}},\ \bibinfo {pages} {753} (\bibinfo {year} {2006})}\BibitemShut
  {NoStop}%
\bibitem [{\citenamefont {Brommer}\ and\ \citenamefont
  {G\"{a}hler}(2007)}]{potfit:2007}%
  \BibitemOpen
  \bibfield  {author} {\bibinfo {author} {\bibfnamefont {P.}~\bibnamefont
  {Brommer}}\ and\ \bibinfo {author} {\bibfnamefont {F.}~\bibnamefont
  {G\"{a}hler}},\ }\href {\doibase 10.1088/0965-0393/15/3/008} {\bibfield
  {journal} {\bibinfo  {journal} {Modelling and Simulation in Materials Science
  and Engineering}\ }\textbf {\bibinfo {volume} {15}},\ \bibinfo {pages} {295}
  (\bibinfo {year} {2007})}\BibitemShut {NoStop}%
\bibitem [{\citenamefont {Brommer}\ \emph {et~al.}(2015)\citenamefont
  {Brommer}, \citenamefont {Kiselev}, \citenamefont {Schopf}, \citenamefont
  {Beck}, \citenamefont {Roth},\ and\ \citenamefont {Trebin}}]{potfit:2015}%
  \BibitemOpen
  \bibfield  {author} {\bibinfo {author} {\bibfnamefont {P.}~\bibnamefont
  {Brommer}}, \bibinfo {author} {\bibfnamefont {A.}~\bibnamefont {Kiselev}},
  \bibinfo {author} {\bibfnamefont {D.}~\bibnamefont {Schopf}}, \bibinfo
  {author} {\bibfnamefont {P.}~\bibnamefont {Beck}}, \bibinfo {author}
  {\bibfnamefont {J.}~\bibnamefont {Roth}}, \ and\ \bibinfo {author}
  {\bibfnamefont {H.-R.}\ \bibnamefont {Trebin}},\ }\href {\doibase
  10.1088/0965-0393/23/7/074002} {\bibfield  {journal} {\bibinfo  {journal}
  {Modelling and Simulation in Materials Science and Engineering}\ }\textbf
  {\bibinfo {volume} {23}},\ \bibinfo {pages} {074002} (\bibinfo {year}
  {2015})}\BibitemShut {NoStop}%
\bibitem [{\citenamefont {Janssen}\ \emph {et~al.}(2019)\citenamefont
  {Janssen}, \citenamefont {Surendralal}, \citenamefont {Lysogorskiy},
  \citenamefont {Todorova}, \citenamefont {Hickel}, \citenamefont {Drautz},\
  and\ \citenamefont {Neugebauer}}]{pyiron}%
  \BibitemOpen
  \bibfield  {author} {\bibinfo {author} {\bibfnamefont {J.}~\bibnamefont
  {Janssen}}, \bibinfo {author} {\bibfnamefont {S.}~\bibnamefont
  {Surendralal}}, \bibinfo {author} {\bibfnamefont {Y.}~\bibnamefont
  {Lysogorskiy}}, \bibinfo {author} {\bibfnamefont {M.}~\bibnamefont
  {Todorova}}, \bibinfo {author} {\bibfnamefont {T.}~\bibnamefont {Hickel}},
  \bibinfo {author} {\bibfnamefont {R.}~\bibnamefont {Drautz}}, \ and\ \bibinfo
  {author} {\bibfnamefont {J.}~\bibnamefont {Neugebauer}},\ }\href {\doibase
  10.1016/j.commatsci.2018.07.043} {\bibfield  {journal} {\bibinfo  {journal}
  {Computational Materials Science}\ }\textbf {\bibinfo {volume} {163}},\
  \bibinfo {pages} {24 } (\bibinfo {year} {2019})}\BibitemShut {NoStop}%
\bibitem [{\citenamefont {Tadmor}, \citenamefont {Ortiz},\ and\ \citenamefont
  {Phillips}(1996)}]{tadmor:ortiz:1996}%
  \BibitemOpen
  \bibfield  {author} {\bibinfo {author} {\bibfnamefont {E.~B.}\ \bibnamefont
  {Tadmor}}, \bibinfo {author} {\bibfnamefont {M.}~\bibnamefont {Ortiz}}, \
  and\ \bibinfo {author} {\bibfnamefont {R.}~\bibnamefont {Phillips}},\ }\href
  {\doibase 10.1080/01418619608243000} {\bibfield  {journal} {\bibinfo
  {journal} {Philosophical Magazine A}\ }\textbf {\bibinfo {volume} {73}},\
  \bibinfo {pages} {1529} (\bibinfo {year} {1996})}\BibitemShut {NoStop}%
\bibitem [{\citenamefont {Tadmor}\ \emph
  {et~al.}(2013{\natexlab{b}})\citenamefont {Tadmor}, \citenamefont {Legoll},
  \citenamefont {Kim}, \citenamefont {Dupuy},\ and\ \citenamefont
  {Miller}}]{tadmor:legoll:2013}%
  \BibitemOpen
  \bibfield  {author} {\bibinfo {author} {\bibfnamefont {E.~B.}\ \bibnamefont
  {Tadmor}}, \bibinfo {author} {\bibfnamefont {F.}~\bibnamefont {Legoll}},
  \bibinfo {author} {\bibfnamefont {W.~K.}\ \bibnamefont {Kim}}, \bibinfo
  {author} {\bibfnamefont {L.~M.}\ \bibnamefont {Dupuy}}, \ and\ \bibinfo
  {author} {\bibfnamefont {R.~E.}\ \bibnamefont {Miller}},\ }\href {\doibase
  10.1115/1.4023013} {\bibfield  {journal} {\bibinfo  {journal} {Applied
  Mechanics Reviews}\ }\textbf {\bibinfo {volume} {65}},\ \bibinfo {pages}
  {010803} (\bibinfo {year} {2013}{\natexlab{b}})}\BibitemShut {NoStop}%
\bibitem [{\citenamefont {Becker}\ \emph {et~al.}(2013)\citenamefont {Becker},
  \citenamefont {Tavazza}, \citenamefont {Trautt},\ and\ \citenamefont
  {de~Macedo}}]{nist_ipr1}%
  \BibitemOpen
  \bibfield  {author} {\bibinfo {author} {\bibfnamefont {C.~A.}\ \bibnamefont
  {Becker}}, \bibinfo {author} {\bibfnamefont {F.}~\bibnamefont {Tavazza}},
  \bibinfo {author} {\bibfnamefont {Z.~T.}\ \bibnamefont {Trautt}}, \ and\
  \bibinfo {author} {\bibfnamefont {R.~A.~B.}\ \bibnamefont {de~Macedo}},\
  }\href {\doibase 10.1016/j.cossms.2013.10.001} {\bibfield  {journal}
  {\bibinfo  {journal} {Current Opinion in Solid State and Materials Science}\
  }\textbf {\bibinfo {volume} {17}},\ \bibinfo {pages} {277 } (\bibinfo {year}
  {2013})},\ \bibinfo {note} {frontiers in Methods for Materials
  Simulations}\BibitemShut {NoStop}%
\bibitem [{\citenamefont {Hale}, \citenamefont {Trautt},\ and\ \citenamefont
  {Becker}(2018)}]{nist_ipr2}%
  \BibitemOpen
  \bibfield  {author} {\bibinfo {author} {\bibfnamefont {L.~M.}\ \bibnamefont
  {Hale}}, \bibinfo {author} {\bibfnamefont {Z.~T.}\ \bibnamefont {Trautt}}, \
  and\ \bibinfo {author} {\bibfnamefont {C.~A.}\ \bibnamefont {Becker}},\
  }\href {\doibase 10.1088/1361-651x/aabc05} {\bibfield  {journal} {\bibinfo
  {journal} {Modelling and Simulation in Materials Science and Engineering}\
  }\textbf {\bibinfo {volume} {26}},\ \bibinfo {pages} {055003} (\bibinfo
  {year} {2018})}\BibitemShut {NoStop}%
\bibitem [{\citenamefont {Choudhary}\ \emph {et~al.}(2017)\citenamefont
  {Choudhary}, \citenamefont {Congo}, \citenamefont {Liang}, \citenamefont
  {Becker}, \citenamefont {Hennig},\ and\ \citenamefont {Tavazza}}]{jarvisff1}%
  \BibitemOpen
  \bibfield  {author} {\bibinfo {author} {\bibfnamefont {K.}~\bibnamefont
  {Choudhary}}, \bibinfo {author} {\bibfnamefont {F.~Y.~P.}\ \bibnamefont
  {Congo}}, \bibinfo {author} {\bibfnamefont {T.}~\bibnamefont {Liang}},
  \bibinfo {author} {\bibfnamefont {C.}~\bibnamefont {Becker}}, \bibinfo
  {author} {\bibfnamefont {R.~G.}\ \bibnamefont {Hennig}}, \ and\ \bibinfo
  {author} {\bibfnamefont {F.}~\bibnamefont {Tavazza}},\ }\href {\doibase
  10.1038/sdata.2016.125} {\bibfield  {journal} {\bibinfo  {journal}
  {Scientific Data}\ }\textbf {\bibinfo {volume} {4}},\ \bibinfo {pages}
  {160125} (\bibinfo {year} {2017})}\BibitemShut {NoStop}%
\bibitem [{\citenamefont {Choudhary}\ \emph {et~al.}(2018)\citenamefont
  {Choudhary}, \citenamefont {Biacchi}, \citenamefont {Ghosh}, \citenamefont
  {Hale}, \citenamefont {Walker},\ and\ \citenamefont {Tavazza}}]{jarvisff2}%
  \BibitemOpen
  \bibfield  {author} {\bibinfo {author} {\bibfnamefont {K.}~\bibnamefont
  {Choudhary}}, \bibinfo {author} {\bibfnamefont {A.~J.}\ \bibnamefont
  {Biacchi}}, \bibinfo {author} {\bibfnamefont {S.}~\bibnamefont {Ghosh}},
  \bibinfo {author} {\bibfnamefont {L.}~\bibnamefont {Hale}}, \bibinfo {author}
  {\bibfnamefont {A.~R.~H.}\ \bibnamefont {Walker}}, \ and\ \bibinfo {author}
  {\bibfnamefont {F.}~\bibnamefont {Tavazza}},\ }\href {\doibase
  10.1088/1361-648x/aadaff} {\bibfield  {journal} {\bibinfo  {journal} {Journal
  of Physics: Condensed Matter}\ }\textbf {\bibinfo {volume} {30}},\ \bibinfo
  {pages} {395901} (\bibinfo {year} {2018})}\BibitemShut {NoStop}%
\bibitem [{\citenamefont {Wen}\ \emph {et~al.}(2015)\citenamefont {Wen},
  \citenamefont {Whalen}, \citenamefont {Elliott},\ and\ \citenamefont
  {Tadmor}}]{wen:whalen:elliott:tadmor:2015}%
  \BibitemOpen
  \bibfield  {author} {\bibinfo {author} {\bibfnamefont {M.}~\bibnamefont
  {Wen}}, \bibinfo {author} {\bibfnamefont {S.~M.}\ \bibnamefont {Whalen}},
  \bibinfo {author} {\bibfnamefont {R.~S.}\ \bibnamefont {Elliott}}, \ and\
  \bibinfo {author} {\bibfnamefont {E.~B.}\ \bibnamefont {Tadmor}},\ }\href
  {\doibase 10.1088/0965-0393/23/7/074008} {\bibfield  {journal} {\bibinfo
  {journal} {Modelling and Simulation in Materials Science and Engineering}\
  }\textbf {\bibinfo {volume} {23}},\ \bibinfo {pages} {074008} (\bibinfo
  {year} {2015})}\BibitemShut {NoStop}%
\bibitem [{\citenamefont {Karls}()}]{kim_query}%
  \BibitemOpen
  \bibfield  {author} {\bibinfo {author} {\bibfnamefont {D.~S.}\ \bibnamefont
  {Karls}},\ }\href@noop {} {}\bibinfo {howpublished}
  {\url{https://github.com/openkim/kim-query}}\BibitemShut {NoStop}%
\bibitem [{\citenamefont {Tadmor}, \citenamefont {Elliott},\ and\ \citenamefont
  {Karls}()}]{kim_property_definition}%
  \BibitemOpen
  \bibfield  {author} {\bibinfo {author} {\bibfnamefont {E.~B.}\ \bibnamefont
  {Tadmor}}, \bibinfo {author} {\bibfnamefont {R.~S.}\ \bibnamefont {Elliott}},
  \ and\ \bibinfo {author} {\bibfnamefont {D.~S.}\ \bibnamefont {Karls}},\
  }\href@noop {} {}\bibinfo {howpublished}
  {\url{https://openkim/properties}}\BibitemShut {NoStop}%
\bibitem [{\citenamefont {Curnow}\ and\ \citenamefont
  {Wichmann}(1976)}]{whetstone}%
  \BibitemOpen
  \bibfield  {author} {\bibinfo {author} {\bibfnamefont {H.~J.}\ \bibnamefont
  {Curnow}}\ and\ \bibinfo {author} {\bibfnamefont {B.~A.}\ \bibnamefont
  {Wichmann}},\ }\href {\doibase 10.1093/comjnl/19.1.43} {\bibfield  {journal}
  {\bibinfo  {journal} {The Computer Journal}\ }\textbf {\bibinfo {volume}
  {19}},\ \bibinfo {pages} {43} (\bibinfo {year} {1976})},\ \Eprint
  {http://arxiv.org/abs/\url{https://academic.oup.com/comjnl/article-pdf/19/1/43/1057793/190043.pdf}}
  {\url{https://academic.oup.com/comjnl/article-pdf/19/1/43/1057793/190043.pdf}}
  \BibitemShut {NoStop}%
\bibitem [{\citenamefont {Daw}, \citenamefont {Foiles},\ and\ \citenamefont
  {Baskes}(1993)}]{daw:foiles:baskes:1993}%
  \BibitemOpen
  \bibfield  {author} {\bibinfo {author} {\bibfnamefont {M.~S.}\ \bibnamefont
  {Daw}}, \bibinfo {author} {\bibfnamefont {S.~M.}\ \bibnamefont {Foiles}}, \
  and\ \bibinfo {author} {\bibfnamefont {M.~I.}\ \bibnamefont {Baskes}},\
  }\href {\doibase http://dx.doi.org/10.1016/0920-2307(93)90001-U} {\bibfield
  {journal} {\bibinfo  {journal} {Materials Science Reports}\ }\textbf
  {\bibinfo {volume} {9}},\ \bibinfo {pages} {251 } (\bibinfo {year}
  {1993})}\BibitemShut {NoStop}%
\bibitem [{\citenamefont {Reddy}(2011)}]{reddy:2011}%
  \BibitemOpen
  \bibfield  {author} {\bibinfo {author} {\bibfnamefont {M.}~\bibnamefont
  {Reddy}},\ }\href@noop {} {\emph {\bibinfo {title} {{API} design for
  {C++}}}},\ \bibinfo {edition} {1st}\ ed.\ (\bibinfo  {publisher} {Morgan
  Kaufmann},\ \bibinfo {address} {Burlington, MA},\ \bibinfo {year}
  {2011})\BibitemShut {NoStop}%
\bibitem [{\citenamefont {Merkel}(2014)}]{docker}%
  \BibitemOpen
  \bibfield  {author} {\bibinfo {author} {\bibfnamefont {D.}~\bibnamefont
  {Merkel}},\ }\href@noop {} {\bibfield  {journal} {\bibinfo  {journal} {Linux
  Journal}\ }\textbf {\bibinfo {volume} {2014}} (\bibinfo {year}
  {2014})}\BibitemShut {NoStop}%
\bibitem [{Note1()}]{Note1}%
  \BibitemOpen
  \bibinfo {note} {For HPC environments, Singularity~\cite {singularity} images
  can be constructed from Docker images.}\BibitemShut {Stop}%
\bibitem [{rsy()}]{rsync}%
  \BibitemOpen
  \href@noop {} {}\bibinfo {howpublished}
  {\url{https://rsync.samba.org}}\BibitemShut {NoStop}%
\bibitem [{tor()}]{tornado}%
  \BibitemOpen
  \href@noop {} {}\bibinfo {howpublished}
  {\url{https://www.tornadoweb.org}}\BibitemShut {NoStop}%
\bibitem [{mon()}]{mongo}%
  \BibitemOpen
  \href@noop {} {}\bibinfo {howpublished}
  {\url{https://www.mongodb.com}}\BibitemShut {NoStop}%
\bibitem [{sql()}]{sqlite}%
  \BibitemOpen
  \href@noop {} {}\bibinfo {howpublished}
  {\url{https://www.sqlite.org}}\BibitemShut {NoStop}%
\bibitem [{\citenamefont {Solem}\ \emph {et~al.}()\citenamefont {Solem} \emph
  {et~al.}}]{celery}%
  \BibitemOpen
  \bibfield  {author} {\bibinfo {author} {\bibfnamefont {A.}~\bibnamefont
  {Solem}} \emph {et~al.},\ }\href@noop {} {\enquote {\bibinfo {title} {Celery
  distributed task queue},}\ }\bibinfo {howpublished}
  {\url{www.celeryproject.org}}\BibitemShut {NoStop}%
\bibitem [{rab()}]{rabbitmq}%
  \BibitemOpen
  \href@noop {} {}\bibinfo {howpublished}
  {\url{https://www.rabbitmq.com}}\BibitemShut {NoStop}%
\bibitem [{Note2()}]{Note2}%
  \BibitemOpen
  \bibinfo {note} {Currently, RabbitMQ features native support only for AMQP
  version 0.9.1, employed here.}\BibitemShut {Stop}%
\bibitem [{Note3()}]{Note3}%
  \BibitemOpen
  \bibinfo {note} {Note that the kim-property python package~\cite
  {kim_property} can be used to create and write property instances. A native
  implementation in LAMMPS is also available.}\BibitemShut {Stop}%
\bibitem [{Note4()}]{Note4}%
  \BibitemOpen
  \bibinfo {note} {Strictly speaking, what is listed are \protect \emph
  {lineages} of {\protect \sf Tests}\protect \xspace , which encompass all
  versions of that {\protect \sf Test}\protect \xspace . The dependency is
  always taken to correspond to the latest existing version in that
  lineage.}\BibitemShut {Stop}%
\bibitem [{Note5()}]{Note5}%
  \BibitemOpen
  \bibinfo {note} {This is applicable in the event where a new version of an
  existing {\protect \sf Test}\protect \xspace is uploaded, which forces its
  downstream dependents to be rerun. The reason is that jobs associated with
  the downstream dependents being removed from the list could otherwise
  eventually be run twice when downstream resolution is performed on the
  {\protect \sf Test~Results}\protect \xspace of jobs associated with the
  others. However, this mechanism can fail if more complicated structures exist
  in the dependency graph. A point of future work is to address this
  shortcoming with a global graph traversal method, e.g.\ a topological sorting
  algorithm, while taking care not to needlessly sequentialize jobs in
  independent branches.}\BibitemShut {Stop}%
\bibitem [{kim()}]{kim_compare}%
  \BibitemOpen
  \href@noop {} {}\bibinfo {howpublished}
  {\url{https://openkim.org/compare}}\BibitemShut {NoStop}%
\bibitem [{\citenamefont {Karls}(2016)}]{karls:2016}%
  \BibitemOpen
  \bibfield  {author} {\bibinfo {author} {\bibfnamefont {D.~S.}\ \bibnamefont
  {Karls}},\ }\emph {\bibinfo {title} {Transferability of Empirical Potentials
  and the Knowledgebase of Interatomic Models}},\ \href
  {\url{http://hdl.handle.net/11299/181747}} {Ph.D. thesis},\ \bibinfo
  {school} {University of Minnesota}, \bibinfo {address} {Minneapolis, MN, USA}
  (\bibinfo {year} {2016})\BibitemShut {NoStop}%
\bibitem [{\citenamefont {Bulatov}(2020)}]{bulatov_privcomm}%
  \BibitemOpen
  \bibfield  {author} {\bibinfo {author} {\bibfnamefont {V.~V.}\ \bibnamefont
  {Bulatov}},\ }\href@noop {} {}\bibinfo {howpublished} {{Private
  Communication}} (\bibinfo {year} {2020})\BibitemShut {NoStop}%
\bibitem [{\citenamefont {Zepeda-Ruiz}\ \emph {et~al.}(2019)\citenamefont
  {Zepeda-Ruiz}, \citenamefont {Stukowski}, \citenamefont {Oppelstrup},
  \citenamefont {Bertin}, \citenamefont {Barton}, \citenamefont {Freitas},\
  and\ \citenamefont {Bulatov}}]{zepeda:2019}%
  \BibitemOpen
  \bibfield  {author} {\bibinfo {author} {\bibfnamefont {L.~A.}\ \bibnamefont
  {Zepeda-Ruiz}}, \bibinfo {author} {\bibfnamefont {A.}~\bibnamefont
  {Stukowski}}, \bibinfo {author} {\bibfnamefont {T.}~\bibnamefont
  {Oppelstrup}}, \bibinfo {author} {\bibfnamefont {N.}~\bibnamefont {Bertin}},
  \bibinfo {author} {\bibfnamefont {N.~R.}\ \bibnamefont {Barton}}, \bibinfo
  {author} {\bibfnamefont {R.}~\bibnamefont {Freitas}}, \ and\ \bibinfo
  {author} {\bibfnamefont {V.~V.}\ \bibnamefont {Bulatov}},\ }\href@noop {}
  {\enquote {\bibinfo {title} {Metal hardening in atomistic detail},}\ }
  (\bibinfo {year} {2019}),\ \Eprint {http://arxiv.org/abs/1909.02030}
  {arXiv:1909.02030 [cond-mat.mtrl-sci]} \BibitemShut {NoStop}%
\bibitem [{\citenamefont {Zepeda-Ruiz}\ \emph {et~al.}(2017)\citenamefont
  {Zepeda-Ruiz}, \citenamefont {Stukowski}, \citenamefont {Oppelstrup},\ and\
  \citenamefont {Bulatov}}]{zepeda:2017}%
  \BibitemOpen
  \bibfield  {author} {\bibinfo {author} {\bibfnamefont {L.~A.}\ \bibnamefont
  {Zepeda-Ruiz}}, \bibinfo {author} {\bibfnamefont {A.}~\bibnamefont
  {Stukowski}}, \bibinfo {author} {\bibfnamefont {T.}~\bibnamefont
  {Oppelstrup}}, \ and\ \bibinfo {author} {\bibfnamefont {V.~V.}\ \bibnamefont
  {Bulatov}},\ }\href {\doibase 10.1038/nature23472} {\bibfield  {journal}
  {\bibinfo  {journal} {Nature}\ }\textbf {\bibinfo {volume} {550}},\ \bibinfo
  {pages} {492} (\bibinfo {year} {2017})}\BibitemShut {NoStop}%
\bibitem [{\citenamefont {Rice}, \citenamefont {Beltz},\ and\ \citenamefont
  {Sun}(1992)}]{rice:1992}%
  \BibitemOpen
  \bibfield  {author} {\bibinfo {author} {\bibfnamefont {J.~R.}\ \bibnamefont
  {Rice}}, \bibinfo {author} {\bibfnamefont {G.~E.}\ \bibnamefont {Beltz}}, \
  and\ \bibinfo {author} {\bibfnamefont {Y.}~\bibnamefont {Sun}},\ }\href@noop
  {} {\bibfield  {journal} {\bibinfo  {journal} {Journal of the Mechanics and
  Physics of Solids}\ }\textbf {\bibinfo {volume} {40}},\ \bibinfo {pages}
  {239} (\bibinfo {year} {1992})}\BibitemShut {NoStop}%
\bibitem [{\citenamefont {Tadmor}\ and\ \citenamefont
  {Hai}(2003)}]{tadmor:hai:2003}%
  \BibitemOpen
  \bibfield  {author} {\bibinfo {author} {\bibfnamefont {E.~B.}\ \bibnamefont
  {Tadmor}}\ and\ \bibinfo {author} {\bibfnamefont {S.}~\bibnamefont {Hai}},\
  }\href {\doibase 10.1016/S0022-5096(03)00005-X} {\bibfield  {journal}
  {\bibinfo  {journal} {Journal of the Mechanics and Physics of Solids}\
  }\textbf {\bibinfo {volume} {51}},\ \bibinfo {pages} {765} (\bibinfo {year}
  {2003})}\BibitemShut {NoStop}%
\bibitem [{\citenamefont {Mishin}(2018)}]{MO_346334655118_005}%
  \BibitemOpen
  \bibfield  {author} {\bibinfo {author} {\bibfnamefont {Y.}~\bibnamefont
  {Mishin}},\ }\href {\doibase 10.25950/bbcadadf} {\enquote {\bibinfo {title}
  {{EAM} potential ({LAMMPS} cubic hermite tabulation) for {C}u developed by
  {M}ishin, {M}ehl and {P}apaconstantopoulos (2001) v005},}\ }\bibinfo
  {howpublished} {OpenKIM, \url{https://doi.org/10.25950/bbcadadf}} (\bibinfo
  {year} {2018})\BibitemShut {NoStop}%
\bibitem [{\citenamefont {Elliott}(2018)}]{MD_120291908751_005}%
  \BibitemOpen
  \bibfield  {author} {\bibinfo {author} {\bibfnamefont {R.~S.}\ \bibnamefont
  {Elliott}},\ }\href {\doibase 10.25950/68defa36} {\enquote {\bibinfo {title}
  {{EAM} {M}odel {D}river for tabulated potentials with cubic {H}ermite spline
  interpolation as used in {LAMMPS} v005},}\ }\bibinfo {howpublished} {OpenKIM,
  \url{https://doi.org/10.25950/bbcadadf}} (\bibinfo {year} {2018})\BibitemShut
  {NoStop}%
\bibitem [{\citenamefont {Mishin}\ \emph {et~al.}(2001)\citenamefont {Mishin},
  \citenamefont {Mehl}, \citenamefont {Papaconstantopoulos}, \citenamefont
  {Voter},\ and\ \citenamefont {Kress}}]{MO_346334655118_005a}%
  \BibitemOpen
  \bibfield  {author} {\bibinfo {author} {\bibfnamefont {Y.}~\bibnamefont
  {Mishin}}, \bibinfo {author} {\bibfnamefont {M.~J.}\ \bibnamefont {Mehl}},
  \bibinfo {author} {\bibfnamefont {D.~A.}\ \bibnamefont
  {Papaconstantopoulos}}, \bibinfo {author} {\bibfnamefont {A.~F.}\
  \bibnamefont {Voter}}, \ and\ \bibinfo {author} {\bibfnamefont {J.~D.}\
  \bibnamefont {Kress}},\ }\href {\doibase 10.1103/PhysRevB.63.224106}
  {\bibfield  {journal} {\bibinfo  {journal} {Physical Review B}\ }\textbf
  {\bibinfo {volume} {63}},\ \bibinfo {pages} {224106} (\bibinfo {year}
  {2001})}\BibitemShut {NoStop}%
\bibitem [{\citenamefont {Sweeney}\ and\ \citenamefont
  {Thain}(2018)}]{containerintegration}%
  \BibitemOpen
  \bibfield  {author} {\bibinfo {author} {\bibfnamefont {K.~M.~D.}\
  \bibnamefont {Sweeney}}\ and\ \bibinfo {author} {\bibfnamefont
  {D.}~\bibnamefont {Thain}},\ }in\ \href {\doibase 10.1145/3217880.3217887}
  {\emph {\bibinfo {booktitle} {Proceedings of the 9th Workshop on Scientific
  Cloud Computing}}},\ \bibinfo {series and number} {ScienceCloud{\'{}}18}\
  (\bibinfo  {publisher} {Association for Computing Machinery},\ \bibinfo
  {address} {New York, NY, USA},\ \bibinfo {year} {2018})\BibitemShut {NoStop}%
\bibitem [{\citenamefont {Kurtzer}, \citenamefont {Sochat},\ and\ \citenamefont
  {Bauer}(2017)}]{singularity}%
  \BibitemOpen
  \bibfield  {author} {\bibinfo {author} {\bibfnamefont {G.~M.}\ \bibnamefont
  {Kurtzer}}, \bibinfo {author} {\bibfnamefont {V.}~\bibnamefont {Sochat}}, \
  and\ \bibinfo {author} {\bibfnamefont {M.~W.}\ \bibnamefont {Bauer}},\ }\href
  {\doibase 10.1371/journal.pone.0177459} {\bibfield  {journal} {\bibinfo
  {journal} {PLOS ONE}\ }\textbf {\bibinfo {volume} {12}},\ \bibinfo {pages}
  {1} (\bibinfo {year} {2017})}\BibitemShut {NoStop}%
\bibitem [{\citenamefont {Afshar}()}]{kim_property}%
  \BibitemOpen
  \bibfield  {author} {\bibinfo {author} {\bibfnamefont {Y.}~\bibnamefont
  {Afshar}},\ }\href@noop {} {}\bibinfo {howpublished}
  {\url{https://github.com/openkim/kim-property}}\BibitemShut {NoStop}%
\end{thebibliography}%

\end{document}